%% file: paper.tex
\documentclass[preprintnumbers,superscriptaddress,nofootinbib,aps,prd,floatfix]{revtex4}
\usepackage{amssymb,amsmath,multirow,graphicx,tabularx,epsfig}
\usepackage{color}
\usepackage{float}
\usepackage{multirow}
\usepackage{subfigure}
\usepackage{hyperref}
\usepackage{comment}
\usepackage{mciteplus}
\usepackage{todonotes}
\usepackage{bigints}
\presetkeys{todonotes}{inline}{}

\include{declare}

\def\beq{\begin{equation}}
\def\eeq{\end{equation}}
\def\beqn{\begin{eqnarray}}
\def\eeqn{\end{eqnarray}}

\def\as{\alpha_{\rm S}}

\def\O#1{{\cal O}(\as^{#1})}

\def\eq#1{Eq.~(\ref{#1})}

\def\beq{\begin{equation}}
\def\eeq{\end{equation}}
\def\beqn{\begin{eqnarray}}
\def\eeqn{\end{eqnarray}}

\def\as{\alpha_{\rm S}}

\def\O#1{{\cal O}(\as^{#1})}

\def\eq#1{Eq.~(\ref{#1})}

\begin{document}

\title{Jet properties at high-multiplicity}

\preprint{IPPP/14/104}
\preprint{DCPT/14/208}
\preprint{MCnet-14-24}

\author{Erik Gerwick}
\email{erik.gerwick@phys.uni-goettingen.de}
\affiliation{II. Physikalisches Institut, Universit\"at G\"ottingen, 
 Germany}

\author{Peter Schichtel}
\email{peter.schichtel@durham.ac.uk}
\affiliation{Institut f\"ur Theoretische Physik, Universit\"at
  Heidelberg, Germany, and}
\affiliation{Institute for Particle Physics Phenomenology, Durham University, UK}

\begin{abstract}
We investigate the behaviour of jets at high-multiplicity 
using analytic techniques.  We consider in detail the 
rates, areas and the intermediate splitting scales as a 
function of the number of jets.  In each case, we are 
able to characterise a general scaling behaviour 
characteristic for QCD processes, which we compare with 
results from the parton shower.  The study of 
these observables potentially offers a very general 
handle in the difficult to describe regime of high-jet 
multiplicity.

\end{abstract}

\maketitle

\tableofcontents

\newpage

\section{Introduction}
\label{sec:intro}

Quarks and gluons produced in hard scattering reactions are measured
experimentally as QCD jets.  Connecting hard scattering to the
hadronic final state is an extended regime of parton evolution, where
a large number of perturbatively well defined splittings produce
particles mostly in the direction of hard partons.  In some cases
these also become jets so that the final state contains an appreciable
number of radiated jets.

The theoretical description of radiated jets is often done via parton
shower simulation~\cite{pythia,herwig, sherpa}.  Though very
successful in some regards, there are two downsides to this.  First,
the parton shower is limited in formal accuracy, and second, features
of the resulting distributions may be opaque from only the individual
components.  While we will say very little about the first point,
fortunately, there is an analytic formalism, the generating
functional, which allows us to address the second.

For jets defined using modern jet-finding algorithms, the generating
functional was constructed first for the Durham algorithm
\cite{Catani:1990rr}.  Jet rates in this algorithm are defined in
terms of a single dimensionless parameter $y_{\text{cut}}$.  However,
jets at hadron colliders are typically clustered using the
generalised-$k_t$ class of jet algorithms, for example the $k_t$
\cite{Ellis:1993tq,Catani:1991hj}, anti-$k_t$~\cite{Cacciari:2008gp}
or Cambridge-Aachen~\cite{CamOrig,CamWobisch} algorithm.  Here jets
are defined in terms of a minimum scale $E_R$ and radius $R$.  A
generating functional valid for these algorithms, and resumming
logarithms of the type $\as \log(E/E_R)\log{(1/R)}$ was introduced in
Ref.~\cite{Gerwick:2012fw}.

A useful feature of the generating functional formalism is that the
multiplicity distribution can be understood to all-orders in
perturbation theory.  An example is the analytic solution for the
average jet multiplicity~\cite{Catani:1991pm,Gerwick:2012fw}.
Similarly, the distribution differentially in the jet multiplicity can
also be computed, and in the case of the Durham algorithm was in
Ref.~\cite{Gerwick:2012hq} for particular kinematic limits.  It was
found that jet rates tend to follow one of two scaling
patterns~\cite{Gerwick:2011tm}, which are most easily understood by
considering the ratio of successive jet rates, $\sigma_{n+1} /
\sigma_n$.  In the large logarithmic limit $\log(1/y_{\text{cut}}) \gg
1$, the distribution is that of a Poisson process and follows
\emph{Poisson scaling}, where $\sigma_{n+1} / \sigma_n \sim 1/(n+1)$.
On the other hand, in the limit $\log(1/y_{\text{cut}}) > 1$ with $\as
\log(1/y_{\text{cut}}) \ll 1$, the distribution becomes geometric
$\sigma_{n+1} / \sigma_n \sim \text{constant}$, which was deemed
\emph{staircase scaling}.  While some insights on scaling were given
in Ref.~\cite{Gerwick:2012fw} for the Gen-$k_t$ class of jet
algorithms, a complete derivation of the multiplicity distributions
was not.

The motivation for this paper is to extend (and generalise) the notion
of scaling in the Gen-$k_t$ class of jet algorithms via the Gen-$k_t$
generating functional.  While the idealised scaling patterns can be
worked out analytically, the distributions coming from parton shower
simulation rarely follow these patterns exactly.  As pointed out in
Ref.~\cite{Gerwick:2012fw}, the discrepancy partially originates from
effects which are not included in the generating functional formalism,
for example kinematics and finite area effects, which disappear only
in the exact limits.

For realistic jet radii ($R=0.3-0.5$), a significant effect pushing
jet multiplicities away from idealised staircase scaling is finite
area considerations.  In order to quantify this we provide in the
second part of this paper a detailed analysis of the area distribution
as a function of the multiplicity.  Keeping with the motivation of the
generating functional formalism, we attempt to describe this
distribution analytically using geometric considerations and the exact
collinear structure of the QCD matrix element.

In the final part of this paper, we take a step towards a more
generalised notion of scaling by considering a different observable,
namely the average $k_t$ splitting scale as a function of the
multiplicity.  After computing some analytic results, we are able to
frame the study of this observable in the context of idealised scaling
patterns, one representing a perfect Poisson process, and the other
for an idealised non-Abelian splitting history.

Our motivation for understanding multiplicity distributions which
undergo scaling is that when these properties are sufficiently
generic, they provide useful handles in difficult QCD environments.
For example, while it would be very challenging to predict exact
properties of a 20 (sub)-jet final state, it is much more feasible to
use the shape of first principle distributions in multiplicity
(comparing to say all 10 - 19 jet events) in order to constrain
whether 20 (sub)-jet events are consistent with the QCD background as
a whole.

A significant amount of recent progress has come from applying
analytic techniques to QCD intensive observables, particularly in the
context of sub-jet studies~\cite{Dasgupta:2013ihk,Dasgupta:2013via}.
In the direction of jet multiplicities and a number of other (sub)-jet
properties, very recently logarithms of the type $\as \log(1/R)$ were
resummed in Ref.  \cite{Dasgupta:2014yra}, which are not considered to
all-orders in the present work.  Further studies on the impact of jet
algorithms on resummation, and especially the appearance of logarithms
in $R$ are given in
Refs.~\cite{Alioli:2013hba,Tackmann:2012bt,Ellis:2010rwa}.

This paper is arranged as follows.  We start off in
sec.\ref{sec:ratios} with a brief review of the Gen-$k_t$ generating
functional and proceed to compute the multiplicity distribution.  We
then attempt to quantify area effects and perform a dedicated
comparison to Montecarlo.  In sec.~\ref{sec:splitting} we explore the
average splitting scales for different multiplicities and splitting
histories.  We offer some discussion on ideas in sec.~\ref{sec:conc}
for the applicability of this work in phenomenological studies, which
are saved for future work.  In the appendix we provide more details on
the rate calculations performed in this work.\bigskip

\section{Jet Ratios from the Gen-$k_T$ generating functional}
\label{sec:ratios}

We start here by reviewing the Gen-$k_t$ generating functional.  A
more in depth description is found in Ref.~\cite{Gerwick:2012fw}.  Our
starting point is the fully exponentiated generating functional (see
app.~\ref{app:evo}) for a parton of flavour $i$. In terms of the
opening angle $\xi = 1-\cos\theta$ between the emitter and emitted
parton, and the energy ratio of the scale evolution $e =E/E_R$, we
have
\begin{align}
  \label{eq:evo}
  \Phi_i(e,\xi) &= u \text{~exp} \left[ \int\limits^\xi_{\xi_R}
    \frac{d\xi'}{\xi'} \int\limits^1_{1/e} dz
    \frac{\alpha_s(ze,\xi')}{2\pi} \sum\limits_{j,k}
      P_{i\rightarrow jk}(z) \left( \frac{ \Phi_j(e,\xi') \Phi_k(
          \mathcal{E}(z),\xi') }{ \Phi_i(e,\xi') } -1 \right) 
  \right] \, ,
\end{align}
where $i,j,l\in\{q,\bar{q},g\}$. The function $\mathcal{E}(z)=ze$
except for $g\rightarrow q\bar{q}$ where it is $e$. The one loop
running coupling given by
\begin{align}
  \label{eq:coupl}
  \alpha_S(ze,\xi') &= \frac{ \pi}{ b_0 \text{log} \frac{ z^2e^2E_R^2\xi'}{
      \Lambda^2}} \, , 
\end{align}
is defined in terms of the coupling at the hard scale.  The $-1$ in
the exponent of~\eq{eq:evo} defines the Sudakov form factor
 \begin{align}
\Delta_i(e,\xi) = \text{~exp} \left[- \int\limits^\xi_{\xi_R}
    \frac{d\xi'}{\xi'} \int\limits^1_{1/e} dz
    \frac{\alpha_s(ze,\xi')}{2\pi} \sum\limits_{j,k}
      P_{i\rightarrow jk}(z) \, ,
  \right]
\end{align}
interpreted as the no-emission probability between scales $E$ and
$E_R$ and angular distances from the hemisphere boundary to the
cut-off set by $\xi_R = 1- \cos R$, defined in the frame of the
emitter.  For jet production in the entire phase space one takes $\xi
= 1-\cos{(\pi/2)} = 1$.  If the observable is related to sub-jets
inside of a larger jet of radius $R_{L}$, the correct upper boundary
is $\xi = 1-\cos{(R_L)} $.

Finally, the $n$-th jet rate is obtained by
differentiating~\eq{eq:evo} with respect to the parameter $u$
\begin{equation}
  \sigma_n = \sigma_0 \left. \frac{1}{n!} \frac{d^n}{du^n} \Phi_i(e,
  \xi) \right|_{u=0}.
\end{equation}
A few notes are in order regarding the approximation of the following
calculations.  Analytic results are often quoted at finite order and
at fixed coupling for clarity, while the full results are numerically
evaluated in the plots where stated.  In general, we will use the
small $R$ approximation so that $1-\cos R \approx R^2/2$ where
necessary.  The logarithms we aim to control are the double-leading
$\as \log(e)\log(\xi/\xi_R)$ in the exponent.\medskip

\subsection{QCD scaling limits}
\label{subs:qcd_scaliing}

Scaling patterns in jet-rates correspond to two idealised statistical
distributions.  The first is Poisson scaling where
\begin{equation}
  \label{posita}
  \frac{ \sigma_{n+1} } { \sigma_{n} } \equiv R_\frac{n+1}{n} \; = \;
  \frac{\bar{n}}{n+1} \, ,
\end{equation}
with $\bar{n}$ the average number of jets. \eq{posita} is
characteristic for the large logarithmically dominated regime of QCD.
Staircase scaling on the other hand corresponds to the geometric (or
fractal) regime of QCD radiation
\begin{equation}
  \label{stairtt}
  R_\frac{n+1}{n} \; = R_0
\end{equation}
where $R_0$ is a constant. The extent to which jet distributions at
ATLAS follow these patterns was studied experimentally in
Ref.~\cite{Aad:2013ysa}. A correspondent pheno study can be found
in~\cite{Englert:2011pq}.  Using high-precision multi-jet NLO
calculations this type of behaviour was investigated
in~\cite{Bern:2014voa,Bern:2014cva}. In the context of BSM searches,
Ref.~\cite{Englert:2011cg} studied staircase scaling for SM
backgrounds opposed to new physics decay jets, while
Ref.~\cite{Hedri:2013pvl} probed the extent of a staircase like
distribution in sub-jet multiplicities.

We will now derive \eq{posita} and~\eq{stairtt} in the Gen-$k_t$
generating functional formalism.

\subsubsection*{Poisson scaling}
In the double-logarithmically dominated regime of QCD we expect the
rates to become a Poisson
process~\cite{Gerwick:2011tm,Gerwick:2012hq}. For the Durham algorithm
this is achieved with a small resolution parameter $y_\text{cut}$. The
generalized-$k_T$ version depends in fact on two scale choices. The
spacial jet resolution $\xi_R$ and the allowed energy range defined by
$E_R (= p_T^\text{min})$. In the limit $1 \ge \xi \gg \xi_R$ the
integral is dominated by the $\xi' \approx \xi_R$ region. Thus we find
\begin{align}
  \label{eq:evo_pois}
  \Phi_i(e,\xi) &= u \text{~exp} \left[ \left( u -1 \right)
    \int\limits^\xi_{\xi_R} \frac{d\xi'}{\xi'} \int\limits^1_{1/e} dz
    \frac{\alpha_s(ze,\xi')}{2\pi} \sum\limits_{j,k} P_{i\rightarrow
      jk}(z) \right]\, ,
\end{align}
which is known to produce Poisson scaling. The limit $1/e\rightarrow
0$ depends on the structure of the splitting kernels. These have poles
of the form $1/z$ which means that the $z\approx 0$ region contributes
most. Therefore we also find the solution in~\eq{eq:evo_pois}.
However the $g\rightarrow q\bar{q}$ splitting drops out as its
splitting kernel is not divergent but goes to zero in this limit. Note
that starting with several hard partons we similarly produce a Poisson
process~\cite{Gerwick:2011tm, Gerwick:2012fw}. However, we expect that
the argument breaks down at some multiplicity around $n\approx
\bar{n}$ so that $R_\frac{n+1}{n} < 1$, where $\bar{n}$ is the rate
parameter of the Poisson process.

\subsubsection*{Staircase scaling}
From the generating functional in the Durham algorithm we know that
the staircase limit is the one opposite to the large double-log limit.
Formally, this regime exists when $\as \log (e) \log(1/\xi_R) \ll 1 $.
Thus we study~\eq{eq:evo} in the $e\rightarrow 1$ limit. To work
within the framework of the generating functional, which resums large
logarithms, we therefore need $ \log(1/\xi_R) \gg 1 $. We also focus
on the pure Yang-Mills case for simplicity, though the arguments are
more general. We Taylor expand the integrand of the generating
functional around $z_0\approx 1$ but with $z<1$. Thus, we can
write~\eq{eq:evo} in the form
\begin{align}
  \label{eq:evo_expand}
  \Phi_g(e,\xi) &= u \text{~exp} \left[ \int\limits^\xi_{\xi_R}
    \frac{d\xi'}{\xi'} \int\limits^1_{1/e} dz
    \frac{\alpha_s(z,\xi')}{2\pi} P_{g\rightarrow gg}(z) \left(
      \Phi_g(e,\xi') + \sum\limits^\infty_{n=1} \frac{(e(z-1))^n}{n!}
      \frac{ d^n \Phi_g(e,\xi')}{ de^n } -1 \right) \right].
\end{align}
Taking only the leading behavior of the $n=1$ term into account we are
able to find a closed solution for $\Phi_g$ (see
app.~\ref{app:solve_stair})
\begin{align}
  \label{eq:gf_closed}
  \Phi_g(e,\xi) &\, = \,\frac{ 1 }{ 1 + \frac{ (1-u) }{ u \Delta_g(e,\xi) }
    - (u-1) \chi(e,\xi) } \, \approx \, \frac{ 1 }{ 1 + \frac{ (1-u) }{ u
      \Delta_g(e,\xi) } } + \frac{ (u-1) \chi(e,\xi) }{ (1 + \frac{
      (1-u) }{ u \Delta_g(e,\xi) })^2 }.
\end{align}
The function $\chi(e,\xi)$ is given in the app.~\ref{app:solve_stair}.
By taking successive $u$ derivatives we find that the jet ratios are
\begin{align}
  \label{eq:staircase}
  R_{\frac{n+1}{n}} &= (1-\Delta_g(e,\xi)) \left[ 1 + \left(
      \frac{(1-\Delta_g(e,\xi))^3}{ \chi(e,\xi)
          \Delta^2_g(e,\xi) } - \frac{1}{\Delta_g(e,\xi)}
        -2
        + (n+1) \right)^{-1} \right].
\end{align}
We can check the large $n$ limit of this formula by taking the
resolved limit $\as L^2 \ll 1$ in which case the term in brackets
behaves like $1 + (1/n)$.  This is in agreement with the fitted form
of the resolved coefficients as derived in
Ref.~\cite{Gerwick:2013haa}.

The result in~\eq{eq:staircase} is very similar to the one found using
the Durham algorithm in~\cite{Gerwick:2012hq}.  At high multiplicity
the ratios converge to constant staircase scaling.  However, we find a
very interesting additional feature: a staircase breaking term
indicating that pure staircase scaling is an asymptotic feature in
$n$.  The breaking term enhances the low multiplicity ratios with
respect to pure staircase scaling. Note that, as mentioned above,
formally we need small $\xi_R$. For finite values we expect $\xi_R$
dependent deviations from~\eq{eq:staircase}.\medskip

\subsection{Inclusion of Area effects}
\label{sec:area}

\begin{figure}[t]
  \centering
  \includegraphics[width=0.3\textwidth]{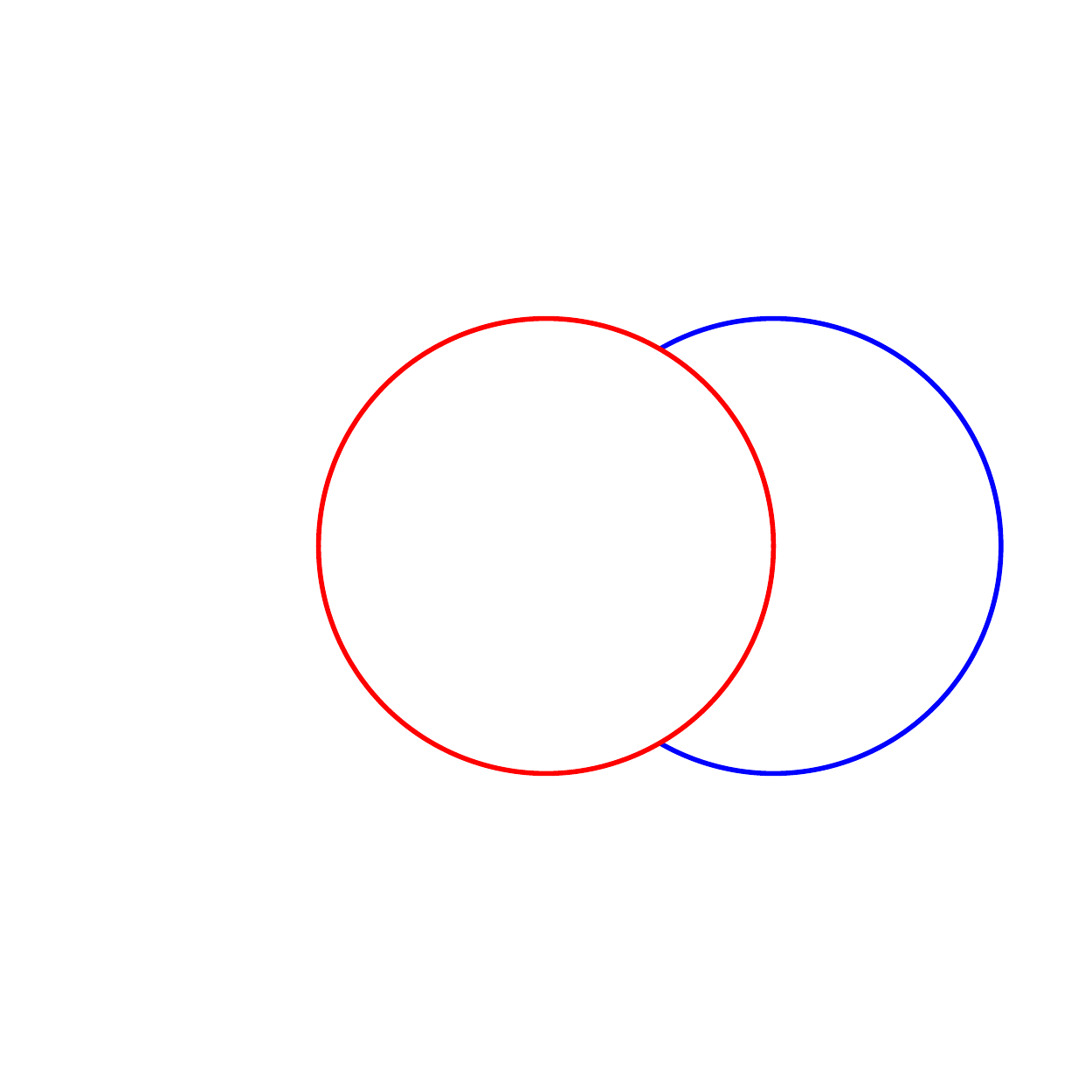}
  \hspace{.5cm}
  \includegraphics[width=0.3\textwidth]{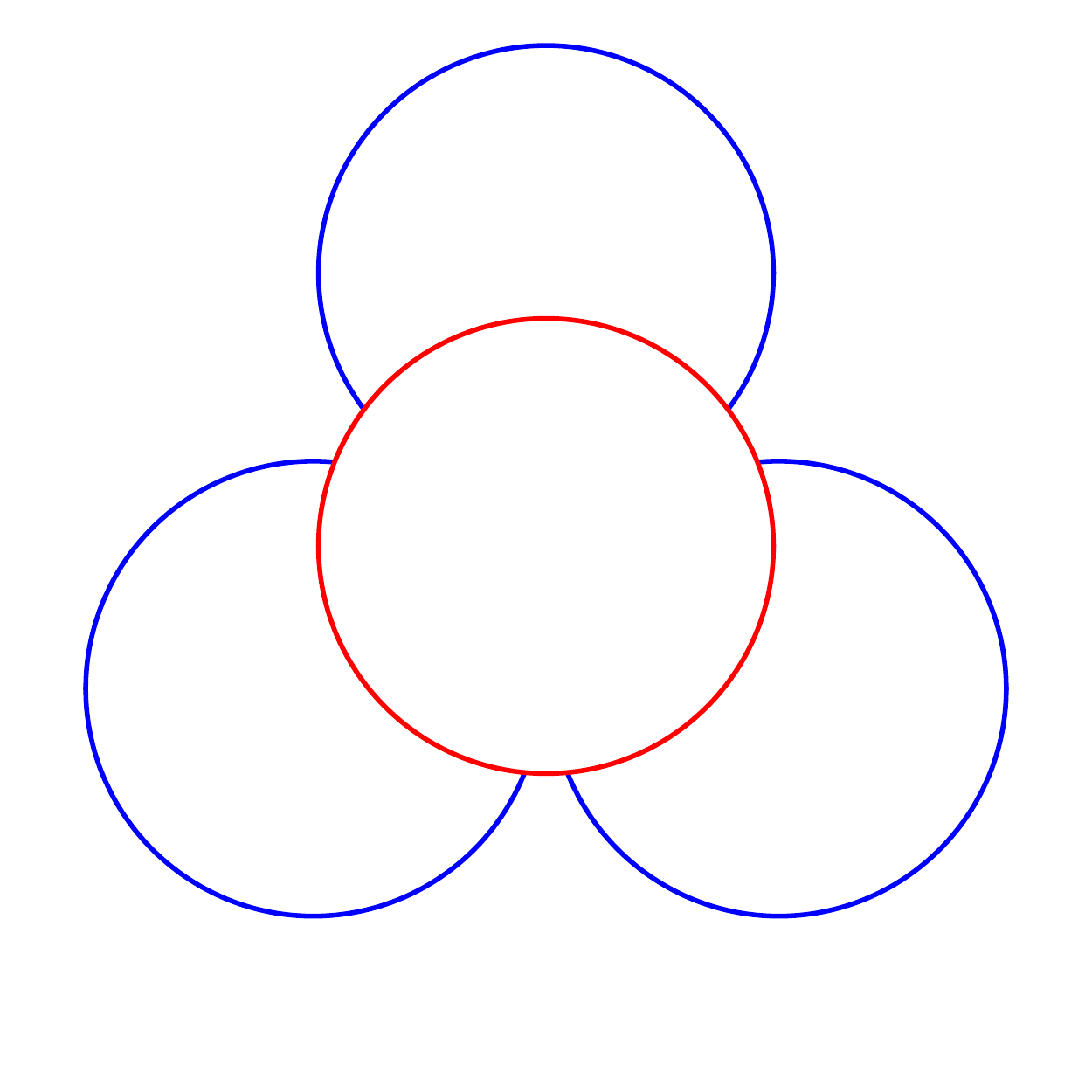}
  \hspace{.5cm}
  \includegraphics[width=0.3\textwidth]{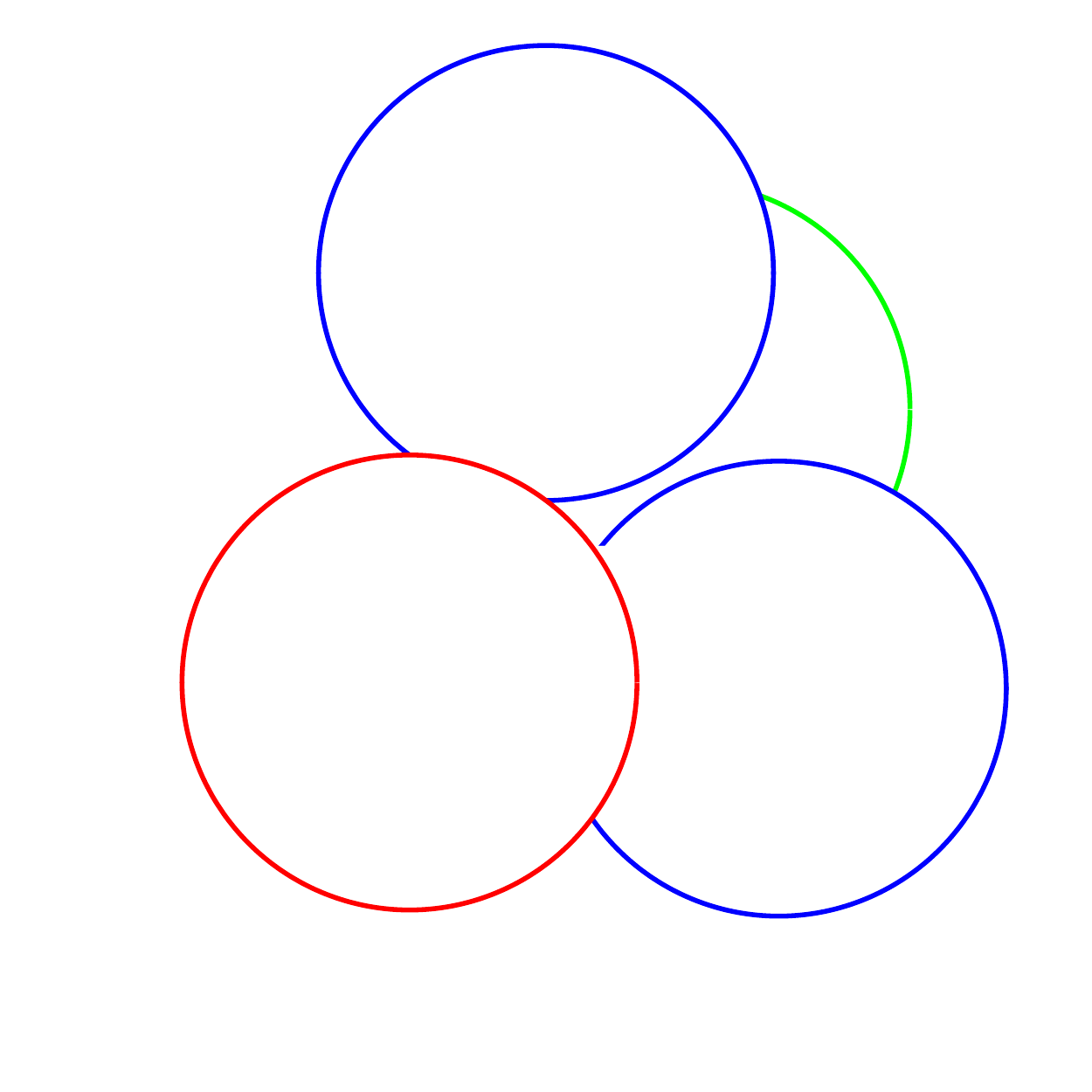}
  \caption {Left: The harder jet (red) defines a disk like object,
    while the softer jet (blue) is partially coverd. This
    defines~\eq{eq:jet_overlap}. Middle: Flower like configuration of
    collinear jets as included in our analytic approach.  Right:
    Non-trivial overlap configuration demonstrating that very small
    jet areas are possible (green).}
  \label{fig:overlapjets}
\end{figure}

\subsubsection*{Naive Phase space effects}

Since we will be studying the high multiplicity behaviour in detail,
an additional effect we would like to include is the simple depletion
of geometric area as the jet multiplicity increases.  The assumptions
we make are that the geometric area of each jet is approximated by
$\pi R^2$ , and that the rates decrease proportional to the available
area uniformly.  This uniquely amounts to an overall suppression
factor at the level of the ratios
\begin{equation}
  \label{eq:naiveps}
  \phi_{\text{ps}} \equiv \frac{\phi(n+1)}{\phi(n)} =
  \frac{1-(n+1)R^2/4}{1-nR^2/4}
\end{equation}
so that the expected behaviour is described by $R_{\frac{n+1}{n}} \to
\phi_{\text{ps}} R_{\frac{n+1}{n}}$. At some $n$ the numerator will
reach $0$ signifying that naively, the entire phase is saturated, and
the ratios go to $0$. Although we will later see that this is an
insufficient approximation it is very useful for some order of
magnitude estimates. We investigate the exact phase space effects in
detail in the next paragraphs.\medskip

\subsubsection*{Evolution equation for jet areas}

In this section, we study how the distribution of jet areas $\rho(A)$
evolves with higher multiplicity.  Jet areas in the generalised-$k_t$
class of jet algorithms was studied first in
Ref.~\cite{Cacciari:2008gn}, with many techniques implemented in
Ref~\cite{Cacciari:2011ma}.  It is a well known feature of the
anti-$k_T$ algorithm that it produces circular
jets~\cite{Cacciari:2008gp}. Furthermore, only with this algorithm is
the area observable IR safe.  We therefore orient this section towards
specifically anti-$k_T$.

Due to the collinear nature of QCD jets quite often overlap, as depict
schematically in the left panel of Fig.~\ref{fig:overlapjets}. This
fact causes the jet areas to follow a non-trivial distribution in
multiplcitiy. We assert that for a fixed number of jets $n$ the
distribution of areas can be parameterised as
\begin{align}
  \label{eq:functional_form_area}
  \rho_n(A) &= \bar{\rho}_{n} \, \delta(A- \bar{A}^\text{full}) + (1-
  \bar{\rho}_{n}) \rho^\text{overlap}_n(A),
\end{align}
where $\bar{A}^\text{full} = 2\pi (1-\text{cos} ( R )) \approx \pi
R^2$ is just the geometric size of a jet of radius $R$ projected on
the sphere and $\bar{\rho}_n$ denotes how many full area jets we
expect, which encodes one part of the non-trivial evolution. The other
non-trivial part is described by $\rho^\text{overlap}_n(A)$, which is
a result of how the jets overlap.

What can we learn about these two quantities?  Let us start with the
simplest case $n=2$, where $\bar{\rho}_2=1$. Therefore, we find
trivially
\begin{align}
  \label{eq:area_distr_2}
  \rho_2(A) &= \delta (A- \bar{A}^\text{full}).
\end{align}
From the collinear approximation we know that jets are distributed due
to $d\xi/\xi$. Assuming fixed coupling and noting that the full area
jet, which is already present, yields a lower bound for the
integration we find
\begin{align}
  \label{eq:rhobar3}
  \bar{\rho}_{3} &= \frac{\text{log} \left(4 \right)-2\, \text{log}
    \left( \text{csc}^2 \left(R\right) \right) }{ \text{log} \left( 1
      - \text{cos} \left(R\right) \right) }.
\end{align}
To find $\rho^\text{overlap}_n(A)$ we need to study how the
geometrical overlap of two jets works. For $n=3$ we can construct that
the non-covered part of the area of a collinear jet which is partially
covered by a full area jet is
\begin{align}
  \label{eq:jet_overlap}
  A^\text{not-covered}(\xi) &= R \left( \sqrt{2\xi- \frac{\xi^2}{2R^2}
    } + 2R\, \text{arcsin} \left( \sqrt{ \frac{\xi}{2R^2} }\right)
  \right) \notag \\
  &\approx \frac{1}{3} \pi R^2 - \sqrt{ \frac{1}{3} } \left(2R^2 -
    R\sqrt{32\,\xi} +\xi \right).
\end{align}
Noting that~\eq{eq:jet_overlap} implies a lower bound of 
\begin{align}
  \label{eq:amin3}
  \bar{A}^\text{min}_{3} &= \frac{1}{6} \left( 3 \sqrt{3} + 2 \pi
  \right) R^2
\end{align}
for the minimal jet size and using again that jets are
distributed due to $d\xi/\xi$ we compute
\begin{align}
  \label{eq:rho3overlap}
  \rho_{3}^\text{overlap}(A) &= \quad\frac{6\, \Theta\left( \left(
    \bar{A}^\text{full} - A\right) \left( A- \bar{A}^\text{min}_{3}
    \right)\right) }{ 6 A - 2 (6 \sqrt{3} + \pi) R^2 + 4 \sqrt{2}\,
    3^\frac{1}{4} R \sqrt{-3 A + \left(6 \sqrt{3} + \pi \right) R^2} }
  \notag\\ &~ \quad \times \frac{1}{ \text{log} \left( 4 + \frac{4
      \pi}{ \sqrt{3}} \right) -2\, \text{arccoth} \left( \frac{ 2\,
      3^\frac{1}{4} }{ \sqrt{ \sqrt{27} - \pi} } \right) }.
\end{align}
In the collinear region this equation is exact. One might be concerned 
that the transition to the $\delta$-function for full area jets is not
smooth. Indeed, if we expand~\eq{eq:jet_overlap} around $\xi=(2R)^2/2$
instead of $R^2/2$ we find a power behaviour with exponent $3/2$. This
causes the measure of the inverse function to diverge. The result is a
divergent, yet integrable, behaviour of $\rho_{3}^\text{overlap}(A
\approx \pi R^2)$. We do not try to merge the two behaviours here,
because this region is shielded by finite resolution effects as we
will see in the following section.

For higher jet multiplicities the situation gets more
complicated. Schematically moving from $n$ to $n+1$ jets is described
by
\begin{align}
  \label{eq:area_evolution}
  \rho_{n+1}(A) &= \frac{1}{n+1} \left( n \, \rho_n(A) +
    \bar{\rho}_{n+1,1} \, \delta \left( A- \bar{A}^\text{full} \right)
    + \left( 1- \bar{\rho}_{n+1,1} \right) \rho_{n+1,1}(A) \right).
\end{align}
Here $1-\bar{\rho}_{n+1,1}$ is how likely it is to add another covered
jet. With $\rho_{n+1,1}(A)$ we describe how the area of this jet is
distributed. For lower multiplicities we might hope to describe both
with their $\rho_3$ equivalents. Deviations from this assumption come
from configurations where the additional jet overlaps with more than
one of the previous ones, which is a configuration we cannot easily
describe with naive geometrical considerations. Of course, from a
certain multiplicity on these configurations will dominate. One of the
features that result from this fact is that we will not have a sharp
drop at $\bar{A}^\text{min}_{3}$ anymore, but a smoother distribution
allowing all possible values for $A$.  In order to estimate at which
value $n$ for the jet multiplicity this will become significant, let
us perform some counting gymnastics for $R=0.5$. Some leading jets
could cover three to four other jets without them touching each
other. Such a configuration would look like a kids version of a
flower, see center panel of Fig.~\ref{fig:overlapjets}. For tighter
configurations we are forced to produce non-trivial overlaps as shown
in the right panel of Fig.~\ref{fig:overlapjets}. Therefore, we should
see a non-trivial structure from starting at $5$ or $6$ jets. However,
choosing this exact configuration is rather unlikely, if we remember
that there is a chance $\mathcal{O}(\bar{\rho}_3)$ to produce full
area jets instead. From the naive phase space considerations,
see~\eq{eq:naiveps}, we know that around $n=15$ the non-overlap
picture must break down. This means that between $n=6$ and $n=15$ the
non-trivial structure becomes significant. The simplest guess would be
that we are able to describe jet area distributions with successive
usage of $\rho_3$ up to $n\approx10$. Using~\eq{eq:area_evolution} we
see that we have $\langle A_n \rangle < \bar{A}^\text{full}$. This
leads to more jets then naively expected. However, generalizing
configurations like in the right panel of Fig.~\ref{fig:overlapjets}
it is clear that this still under-estimates the maximum number of
possible jets. The end point of the spectrum is non-trivial and we
will study it in more detail in the next section.

\subsubsection*{Jet area distribution from MC}

\begin{figure}[t]
  \centering
  \includegraphics[width=0.3\textwidth]{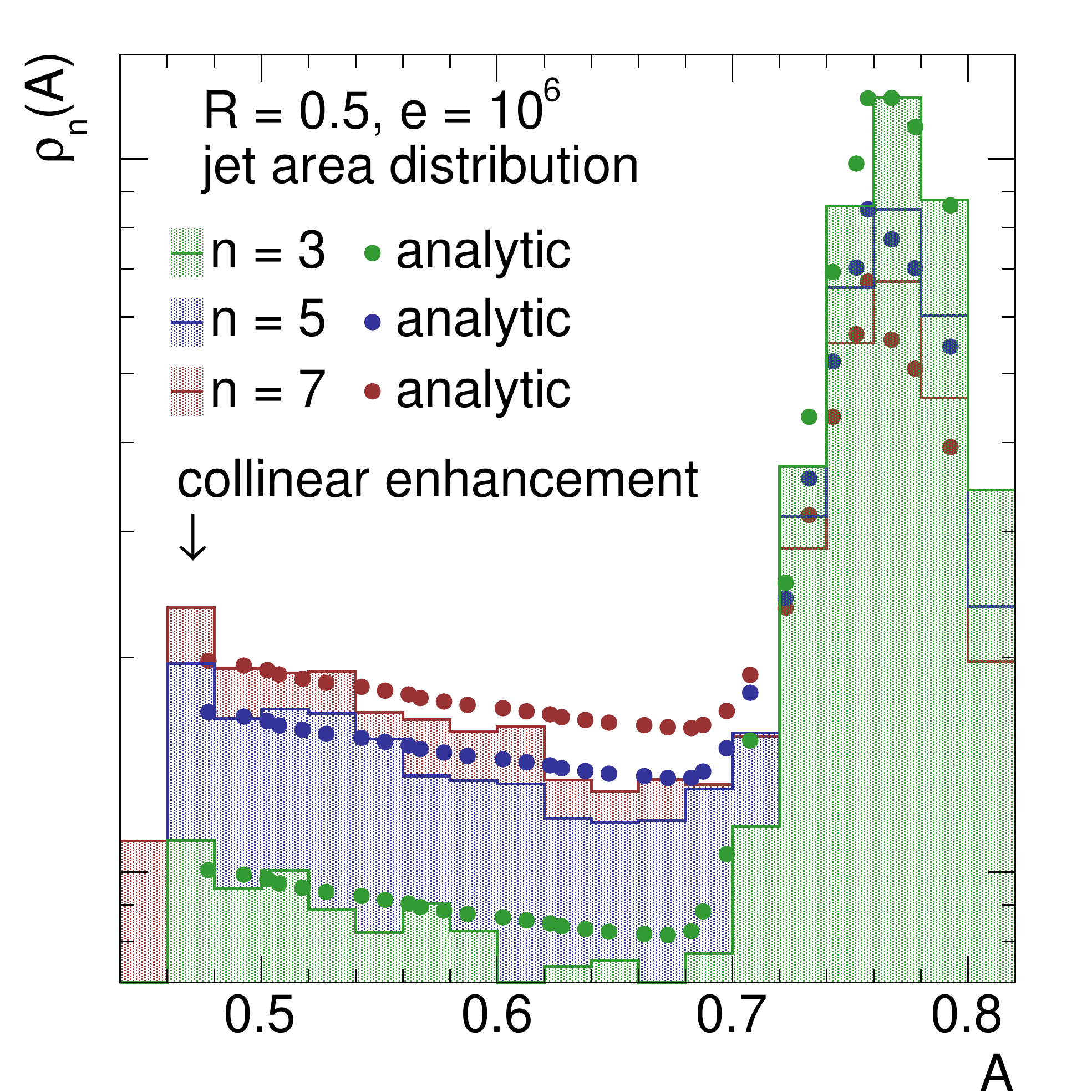}
  \hspace{.5cm}
  \includegraphics[width=0.3\textwidth]{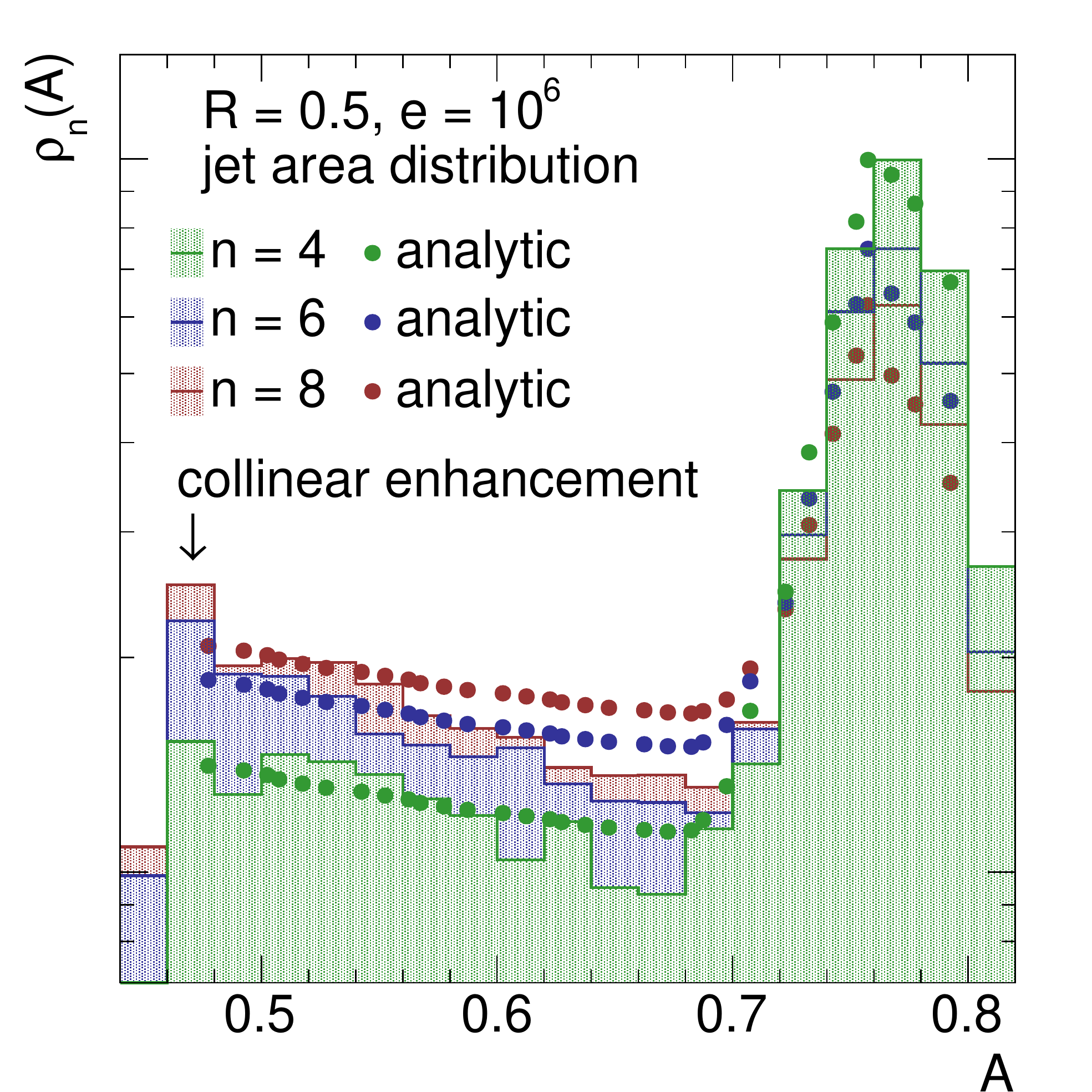}
  \hspace{.5cm}
  \includegraphics[width=0.3\textwidth]{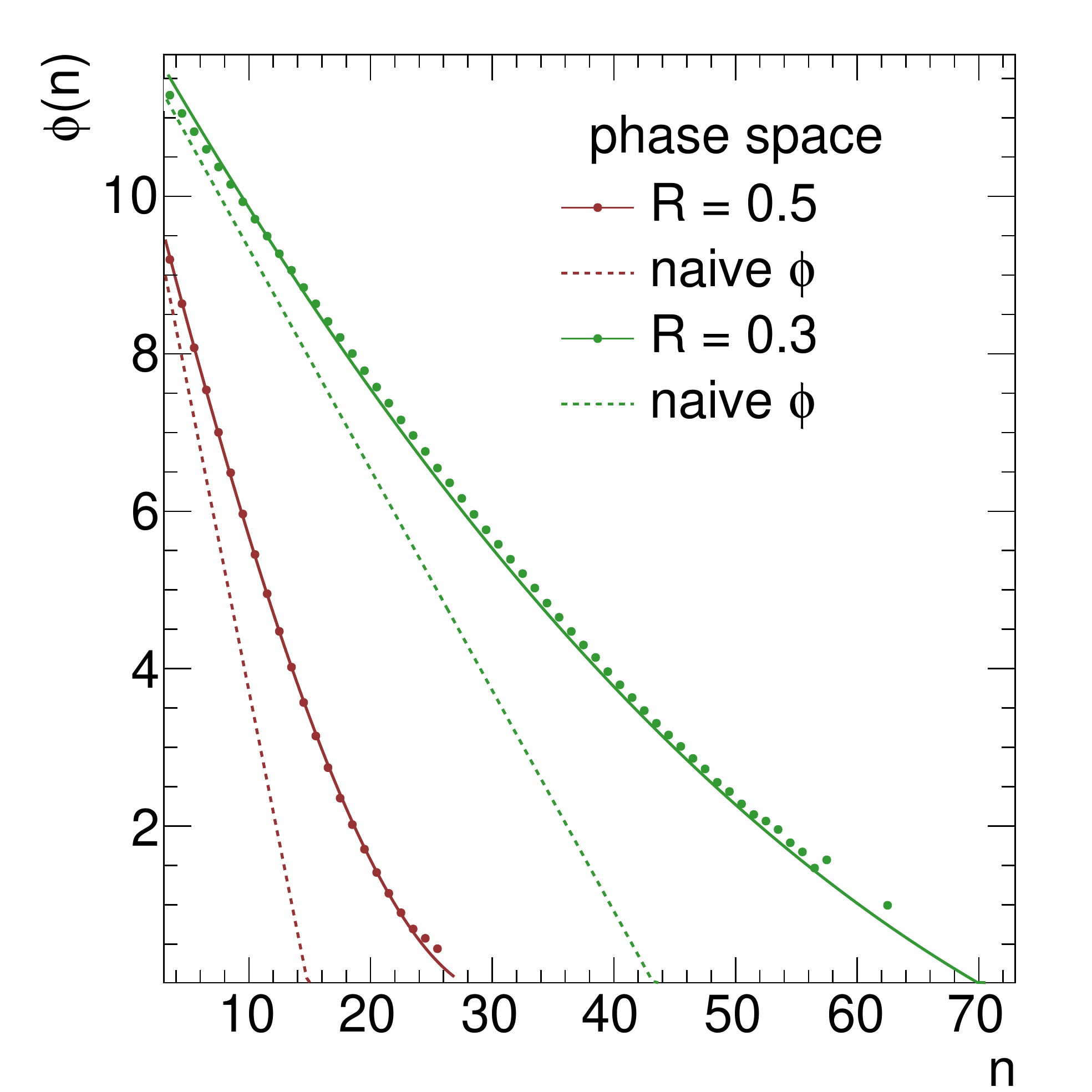}
  \caption{Left and middle panel: Comparison of $\rho_n(A)$
    from~\eq{eq:functional_form_area} (dots) with MC data for $R=0.5$
    from $n=3$ up to $n=8$ (shaded)xs. Right panel: Comparison of the
    actual phase space computed via $\langle A(n) \rangle$ (dots) with
    the naive formula~\eq{eq:naiveps} (dashed) and a polynomial of
    third order (solid) for $R=0.5$ (red) and $R=0.3$ (green).}
  \label{fig:areas}
\end{figure}

In this paragraph we would like to compare our analytic considerations
with a parton shower simulation (specifically the CS shower in
SHERPA~\cite{Gleisberg:2008ta,Schumann:2007mg}) in order to assess
non-trivial behaviour at high multiplicities. For this purpose it is
instructive to recall how jet areas are measured. In the Fastjet
implementation~\cite{Cacciari:2011ma,Cacciari:2008gn} jet areas can be
computed in one of three ways, active area, Passive areas and Veronoi
areas.  We use the active area option of Fastjet, but implement
spherical coordinates.  The active area option divides the
$(\eta,\phi)$-plane into cells of fixed size and places a ghost
particle (almost vanishing $p_T$) randomly in each cell. If the ghost
happens to end up in a certain jet, we count the cell's size towards
that jet's active area.

We study jets of size $R=0.5$ and $R=0.3$. We will also consider jets
of size $R=0.1$ in the next section. However, we will not study them
here, simply because even from naive phase space considerations we
only expect effects after $n\approx 400$. Let us estimate what we
expect for the area measurement. To compensate for the flat
approximation we use an effective radius which corresponds to an area
of $\bar{A}^\text{full}$. Using a random sample of ghosts and jet
midpoints we expect that the area is normal distributed around
$A=0.77$ with a width of $\Delta A = 0.02$ for $R=0.5$ jets and
$A=0.28$ with a width of $\Delta A = 0.016$ for $R=0.3$ jets. The
variance $\Delta A$ is driven by the cell size and the actual jet
radius.

We compare these the analytic formula from the previous section with
MC data, using the numbers in the previous paragraph. For the average
jet size and width we find good agreement. However, the actual width
is slightly larger due to spherical geometry. In Fig.~\ref{fig:areas}
we show the non-trivial area distributions for the exclusive $n=3$ up
to $n=8$ cases for $R=0.5$.  We find good agreement for the three jet
case with~\eq{eq:rho3overlap}. For higher multiplicities we fix the
overall normalisation of~\eq{eq:area_evolution} to produce the same
maximum height as the peak at $\bar{A}^\text{full}$ in the
MC\footnote{This is neccessary because we neglect any possibility of
  non-trivial overlap for our analytic ansatz. This directly leads to
  a different normalization for the two quantities we wish to compare,
  which we compensate with this prescription.}. We observe that we
describe the area overlap distribution accurately for low
multiplicities. As expected, for higher multiplicities $n\approx
\mathcal{O}(10)$ our description breaks down. In the right panel of
Fig.~\ref{fig:areas} we show the phase space for all multiplicities
for $R=0.5$ and $R=0.3$ together with~\eq{eq:naiveps} for the naive
phase space expectation. For that purpose we compute $\phi(n)=4\pi-
\langle A(n) \rangle$. The true phase space follows a polynomial of
third order, where the linear part is driven
by~\eq{eq:rho3overlap}. The higher order terms encode the non-trivial
overlap configurations not accessable with our ansatz.\medskip

\subsection{The full picture}
\label{subs:combine}

\begin{figure}[t]
\includegraphics[width=0.3\textwidth]{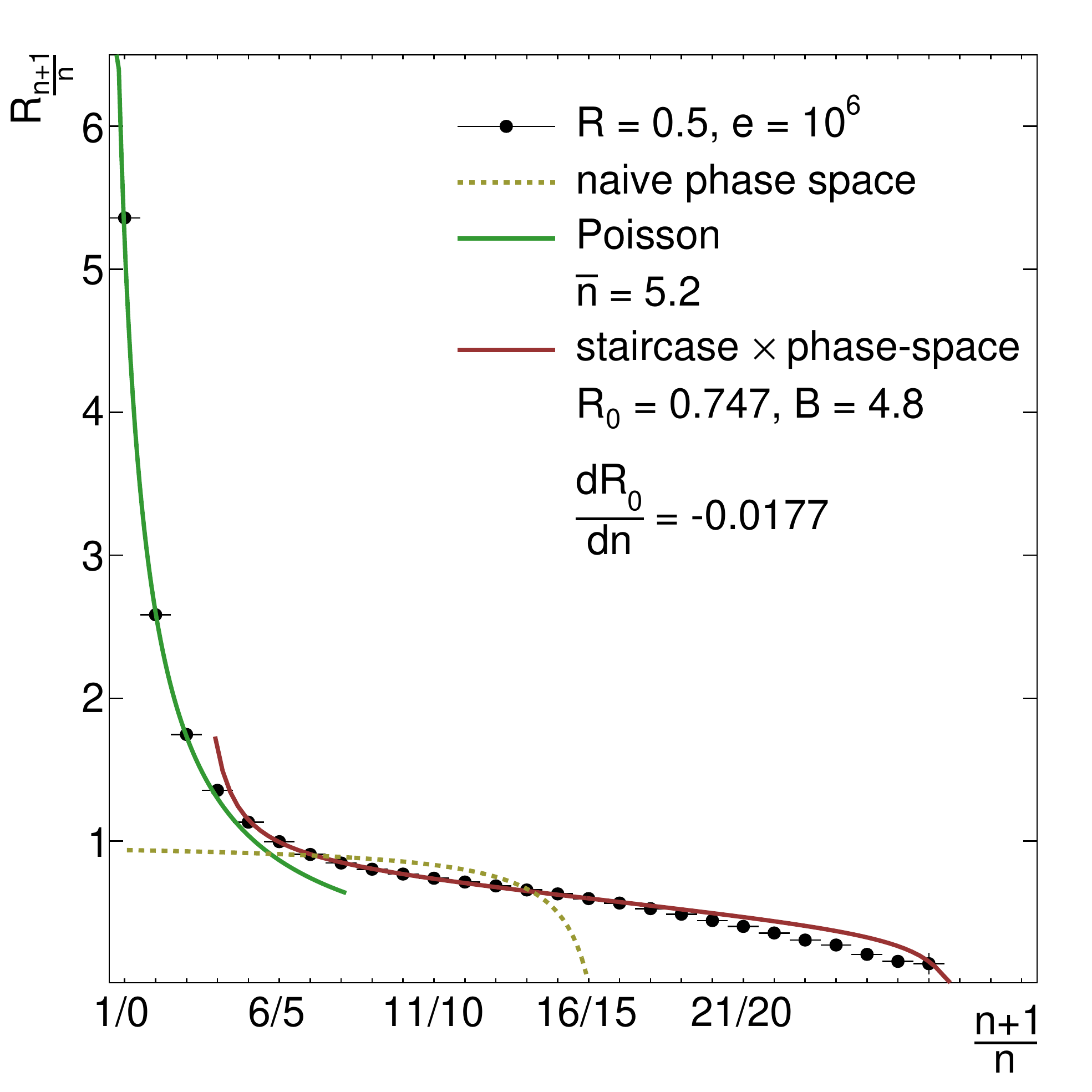}
\hspace{.5cm}
\includegraphics[width=0.3\textwidth]{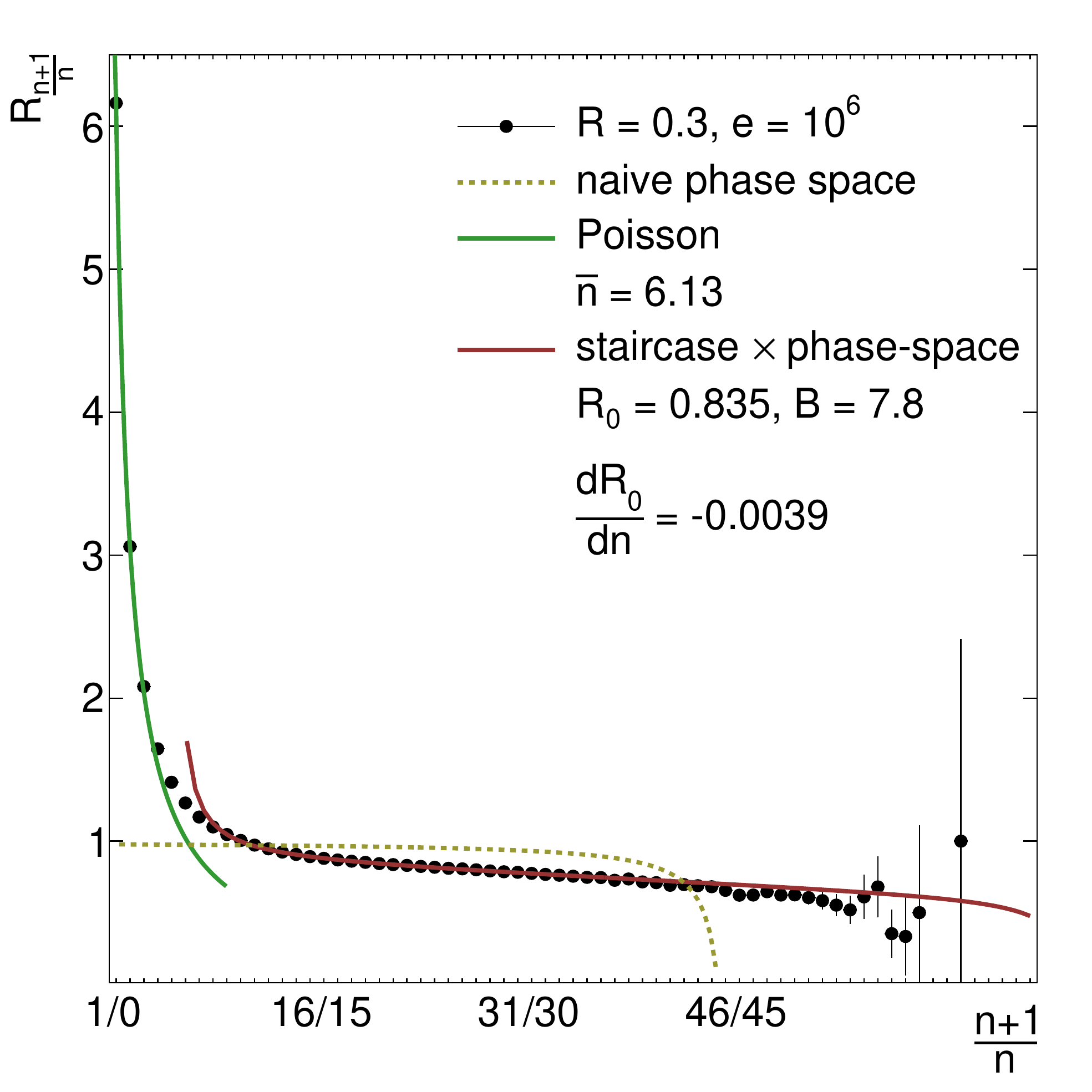}
\hspace{.5cm}
\includegraphics[width=0.3\textwidth]{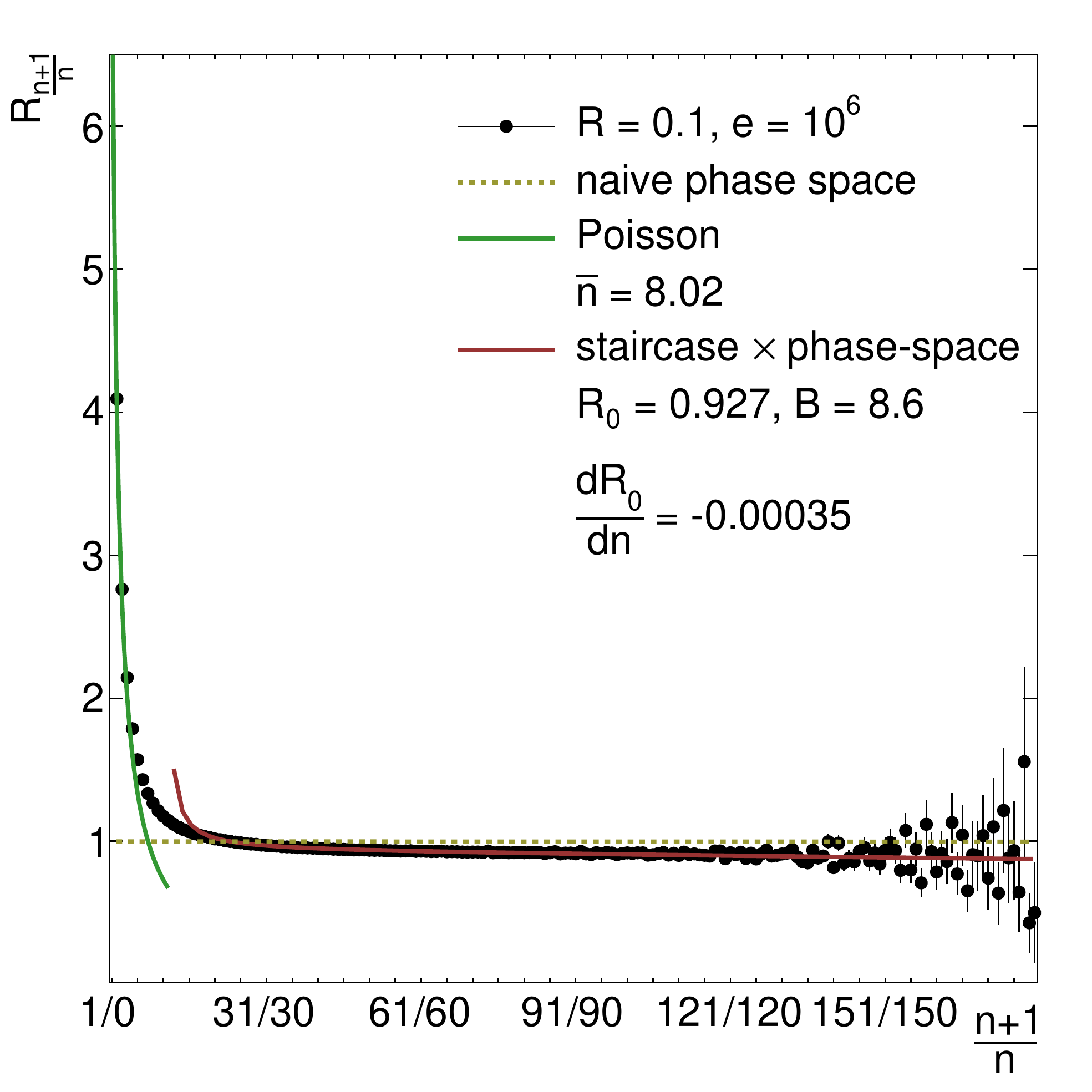}
\caption{Exclusive jet cross section ratios for $R=0.5$ (left),
  $R=0.3$ (middle), and $R=0.1$ (right). The red solid lines show the
  staircase scaling hypothesis times phase space. An additional term
  $dR_0/dn (n+1)$ is fitted to quantify the deviation
  from~\eq{eq:staircase}. In addition we show the naive phase space
  assumption (dashed yellow) and the Poisson scaling hypothesis
  (green).}
\label{fig:ratios}
\end{figure}

Now we would like to put all the different parts together, and see to
what extent we can understand the effects driving the jet ratio
distribution.  In order to maximise statistics we choose a very large
energy hierarchy $E = 10^6 \;\gev$ and $E_R = 1 \gev$, with jets
clustered in the anti-$k_t$ jet algorithm.

We expect low multiplicities to be described by a Poisson process,
defined by $\bar{n} = | \log(\Delta(e,\xi))|$~\cite{Gerwick:2012hq}.
In addition the average number of jets is rather sensitive to the
chosen coupling at the hard scale, which we take as $\alpha_S(E) =
0.076$. We show a comparison in Tab.~\ref{tab:finite_R} and find good
agreement in the double logaritmic region. Finite $R$ effects are
expected from the generating functional as well as the fact that very
low jet multiplicities are matched to the matrix element. At
multiplicities $n \gg \bar{n}$ we have~\eq{eq:staircase}. In between,
there is a somewhat awkward intermediate regime which we cannot say
much about. However, for realistic jet radii $\mathcal{O}(0.5)$ this
regime is very small.  For high multiplicities the pure staircase
scaling is suppressed by phase space effects driven by depletion of
available area as shown in Fig.~\ref{fig:areas}. Note, however, that
for finite $R$ we expect additional effects, which are not included in
our approach. These should vanish as $R\rightarrow 0$.

To demonstrate these claims we show exclusive jet cross section ratios
for $R=0.5$, $R=0.3$ and $R=0.1$ with $e=10^{6}$ in
Fig.~\ref{fig:ratios}. The green line corresponds to the Poisson
hypothesis, while the yellow dashed line
specifies~\eq{eq:naiveps}. The red line is charecterized by
(see~\eq{eq:staircase}),
\begin{align}
  \label{eq:datacomp}
  R_{\frac{n+1}{n}} &= \left( R_0 \left[ 1 + \frac{1}{B+(n+1)} \right]
    + \frac{dR_0}{ dn} (n+1) \right)\times \frac{ \phi(n+1)}{
    \phi(n)},
\end{align}
where the numerical values for $R_0$, $B$, and $dR_0/dn$ are given in
the plots. We take the phase space factor from the polynomial in the
previous section. Note, while there is a deeper connection between
$R_0$ and $B$, as well as there exact position~\eq{eq:datacomp} that
is not true for the term $dR_0/dn$. This term is purely
phenomenological and we introduce it to describe an additional small
tilt we observe for jet cross section ratios. Therefor it quantifies
the discrepancy from our staircase scaling formula. Recall that we
formally need small $R$ for the derivation of~\eq{eq:staircase}. In
Tab.~\ref{tab:finite_R} we show that the fitted deviation $dR_0/dn$ is
small and indeed vanishes as $R\rightarrow 0$.\bigskip

\begin{table}[t]
  \renewcommand\arraystretch{1.4}
  \renewcommand\tabcolsep{5pt}
  \centering
  \begin{tabular}{|r|ccccccccccc|}
    \hline
    $R$ & $0.5$  & $0.4$  & $0.3$  & $0.2$  & $0.1$  & $0.09$  & $0.08$  & $0.07$  & $0.06$  & $0.05$  & $0.04$ \\
    \hline
    \hline
      $|\log(\Delta(e,\xi))|$ & $3.2$  & $3.8$ & $4.7$ & $6.0$ & $8.3$ & $8.7$ & $9.1$ & $9.6$ & $10.1$ & $10.7$& $11.5$ \\
    $\bar{n}$                 & $5.2$  & $5.6$ & $6.1$ & $6.9$ & $8.0$ & $8.9$ & $9.2$ & $9.7$ & $10.0$ & $10.4$& $11.1$ \\
    \hline
    $R_0$                            & $0.75$  & $0.79$ & $0.82$ & $0.857$ & $0.921$ & $0.934$ & $0.937$ & $0.938$ & $0.941$ & $0.945$& $0.953$ \\
    $\left| \frac{dR_0}{dn} \right|$ & $0.018$  & $0.010$ & $0.0039$ & $0.0011$ & $0.00035$ & $0.0003$ & $0.0002$ & $0.0002$ & $0.00013$ & $0.00011$& $0.000074$ \\
    \hline
  \end{tabular}
  \caption{We summarize the relevant quantities for studying
    deviations as a function of $R$. The first two rows show the
    expected and observered Poisson parameter $\bar{n}$. The last two
    rows contain the numerical value for $R_0$ compared to $dR_0/dn$.}
  \label{tab:finite_R}
\end{table}

\section{Distribution of $k_t$ splitting scales}
\label{sec:splitting}

Intermediate splitting scales in multi-jet events are useful primarily
due to the close correspondence with parton shower splitting
variables.  They have been measured extensively in collider
experiments, in $e^+ e^-$~\cite{Pfeifenschneider:1999rz} and more
recently at the LHC in the context of ME-PS
matching~\cite{Aad:2013ueu}.  As our emphasis in this work is scaling
features especially at high-multiplicity, we explore in this section
the distribution of splitting scales as a function of multiplicity
very much in analogy to the jet rates.  The hope is that the resulting
distribution could provide generic first principle handles on QCD
showered events, in connection to what (Staircase/Poisson) scaling
behaviour provides for the rates.

\subsection{General properties of splitting scales}

In order to discuss splitting scales as derived from the 
Gen-$k_t$ generating functional, we first note that 
emission are separately ordered in energy and angle in this 
framework, and not explicitly in $k_t$.  We can 
enforce $k_t$ ordering by hand, with a $\Theta$ function 
for example, bearing in mind that $z E \sqrt{\xi/\xi_R}$ 
corresponds to the usual definition of the splitting scale 
for a $k_t$ jet algorithm~\cite{Catani:1993hr, Ellis:1993tq} in 
the small-$R$ limit.  

Our notation for the splitting scales is that 
$\langle k_t^{(i,j)} \rangle$ is the $j$ hardest 
emission in $k_t$ in a $i$ jet splitting history, so that we 
will always speak of the distribution or ordered scales for 
an exclusive event.   Now let us discuss briefly what we 
expect the resulting distributions to convey regarding 
the nature of QCD showering.

First of all, since we know that the Pseudo-abelian limit 
where all emissions are primary corresponds to a 
Poisson process, the average intermediate scale 
$\langle k_t^{(i)}\rangle$ is independent of the multiplicity.  
This property is true both at the inclusive and exclusive level, 
indicating that it persists for fixed-order and resummed 
calculations.  A way to see that this is 
that the splitting scales correspond to the inter-arrival times 
of the Poisson process, which are themselves memory-less.  

The constant average splitting scales in the pseudo-Abelian limit is
similar in origin to the idealised $\sigma_{n+1} /\sigma_n = 1/(n+1)$
behaviour in the rates for a Poisson process.  In QCD, once we allow
for correlated emissions, the picture will change, and we may ask the
question whether there is some notion of a correlated emission
dominated phase imprinted in the splitting scales.  As we will see,
our computations indicate that there is some evidence for behaviour
like this.

\subsubsection*{One emission}
We start with the single emission splitting scale.  This is the
average value of $E z \sqrt{\xi'/\xi_R}$ of the probability function
$(z\xi')^{-1} \as(zE\xi')\,\Delta(z,\xi')$ in the plane defined by
boundaries $\xi' \in [\xi_R,\xi]$ and $z \in [1/e,1]$.  The
expectation value of the splitting scale $\langle k_t \rangle = z E
\sqrt{\xi'/\xi_R}$ inside the integral is
\begin{alignat}{5}
  \label{1startkt}
  \langle k_t^{(1,1)}\rangle = \dfrac{E \bigintss_{\xi_R}^{\xi}
    \dfrac{d \xi'}{\sqrt{\xi'}} \bigintss_{\,1/e}^{1} d z\, z \, P(z)
    \, \as (zE\xi') \,\Delta(z,\xi')} {
    \sqrt{\xi_R}\bigintss_{\xi_R}^{\xi} \dfrac{d \xi'}{\xi'}
    \bigintss_{\, 1/e}^{1} d z\, P(z) \, \as(zE\xi')\,\Delta(z,\xi') }
\end{alignat}
which keeping only the most singular contributions gives the leading
logarithms
\begin{equation}
\langle k_t^{(1,1)} \rangle = \frac{2\, (E-E_R)(\sqrt{\xi/\xi_R}-1)}
{\log(E/E_R)\log(\sqrt{\xi / \xi_R)}} + 
\O{}
\label{1lowor}
\end{equation}
The expression in~\eq{1startkt} is well approximated by~\eq{1lowor}
for all but large energy scale ratios $\mathcal{O}(10^6)$, or
extremely small jet radii.  This effect can be seen by noting that the
inclusion of higher order terms in the Sudakov affects the numerator
and denominator of~\eq{1startkt} in the same direction, thus leaving a
diminished residual dependence.  Furthermore, there is no leading
order dependence on $\as$, a consequence is that running coupling
effects are pushed to third-order.

\subsubsection*{Two emissions}
In order to compute the average splitting scales for two emission we
distinguish between $k_t^{(21)}$ and $k_t^{(22)}$ such that we have
$k_t^{(21)}> k_t^{(22)} $.  Furthermore, at this multiplicity there
are two splitting histories which we denote correlated and
uncorrelated
\begin{equation*}
\text{uncorrelated\;:} \qquad
\includegraphics[width=0.1\textwidth]{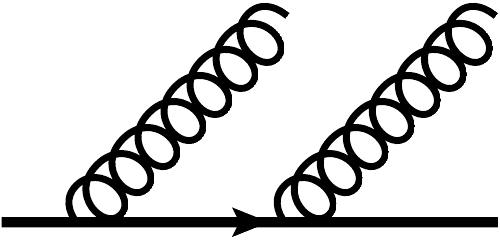} \qquad \qquad
\text{correlated\;:} \qquad
\includegraphics[width=0.1\textwidth]{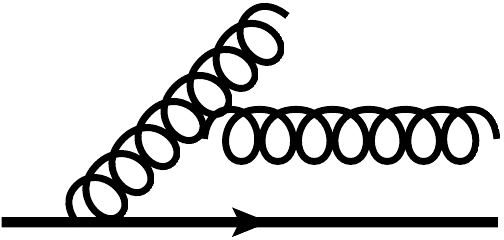} \qquad
\end{equation*}
We write the two primary emission rate in a $k_t$ ordered way such
that the contribution to lowest order is
\begin{alignat}{5}
P_{2}^{(\text{uncorr})} \;  = \; \frac{1}{2} \frac{\as\;  C_F^2}{\pi}& \left( \int_{\xi_R}^{\xi}\frac{d \xi'}{\xi'} 
\int_{E_R/E}^1 dz \, P(z)
\int_{\xi_R}^{\xi}\frac{d \xi''}{\xi''} 
\int_{E_R/E}^1 dz' \, P(z')\; \Theta(z'\sqrt{\xi''} < z\sqrt{\xi'})\right. 
\notag \\
&\left.
+\int_{\xi_R}^{\xi}\frac{d \xi'}{\xi'} 
\int_{E_R/E}^1 dz \, P(z)
\int_{\xi'}^{\xi}\frac{d \xi''}{\xi''} 
\int_{E_R/E}^1 dz' \, P(z')\;
\Theta(z'\sqrt{\xi''} \ge z\sqrt{\xi'})\right)
\label{two_uncorr}
\end{alignat}
which gives the well known result $(1/2!)P_1^2$ for the rate.  The
contributions in~\eq{two_uncorr} arise when the second emission in the
integral is smaller or larger in $k_t$ respectively.  For correlated
emissions, due to ordering only the first term is present.

The $\Theta$ function provides a complicated phase space constraint,
preventing a simple analytic evaluation.  In the inclusive $k_t$
(Durham) algorithm, the $k_t$ regions would factorise, and we would
expect a much simpler final result.  However, in this work we have
stayed as close as possible to the more typically used jet variable
$E_R$ and $\xi_R$.  Despite this complication, we can extract the
average splitting scale of the harder emission
\begin{alignat}{5}
\langle k_t^{(21)} \rangle_{\text{uncorr.}}
= \frac{E}{P_{2}^{(\text{uncorr})} } \frac{\as\;  C_F^2}{2\pi}& \left( \int_{\xi_R}^{\xi}\frac{d \xi'}{\xi'} 
\int_{E_R/E}^1 dz \, P(z) \, z \, \sqrt{\xi'} 
\int_{\xi_R}^{\xi}\frac{d \xi''}{\xi''} 
\int_{E_R/E}^1 dz' \, P(z') \; \Theta(z'\sqrt{\xi''} < z\sqrt{\xi'})\right. 
\notag \\
&\left.
+\int_{\xi_R}^{\xi}\frac{d \xi'}{\xi'} 
\int_{E_R/E}^1 dz \, P(z)
\int_{\xi'}^{\xi}\frac{d \xi''}{\xi''} 
\int_{E_R/E}^1 dz' \, P(z') \, z' \, \sqrt{\xi''}\;
\Theta(z'\sqrt{\xi''} > z\sqrt{\xi'})\right)\, ,
\label{ave_two_uncorr}
\end{alignat}
and similarly for the softer
\begin{alignat}{5}
\langle k_t^{(22)} \rangle_{\text{uncorr.}}
= \frac{E}{P_{2}^{(\text{uncorr})} } \frac{\as\;  C_F^2}{2\pi}& \left( \int_{\xi_R}^{\xi}\frac{d \xi'}{\xi'} 
\int_{E_R/E}^1 dz \, P(z) \,  
\int_{\xi_R}^{\xi}\frac{d \xi''}{\xi''} 
\int_{E_R/E}^1 dz' \, P(z') \, z' \, \sqrt{\xi''}\,  \Theta(z'\sqrt{\xi''} < z\sqrt{\xi'})\right. 
\notag \\
&\left.
+\int_{\xi_R}^{\xi}\frac{d \xi'}{\xi'} 
\int_{E_R/E}^1 dz \, P(z) \ z\, \sqrt{\xi'} 
\int_{\xi'}^{\xi}\frac{d \xi''}{\xi''} 
\int_{E_R/E}^1 dz' \, P(z') \, \;
\Theta(z'\sqrt{\xi''} > z\sqrt{\xi'})\right)\, ,
\label{ave_two_uncorr}
\end{alignat}
We can evaluate these analytically by taking only the energy ordering
enforced by the $\Theta$ functions, which differs from the full result
by sub-leading terms, and demonstrates the same essential behaviour.
In this case we obtain
\begin{alignat}{5}
& \langle k_t^{(21)} \rangle_{\text{uncorr.}} =
\frac{4}{\log^2(E/E_R)\log(\xi/\xi_R)}
\left(E_R-E + E \log\left[ \frac{E}{E_R}\right]\right) 
(\sqrt{\xi/\xi_R}-1),
\notag 
\\
& \langle k_t^{(22)} \rangle_{\text{uncorr.}} 
=
\langle k_t^{(21)} \rangle_{\text{uncorr.}} 
\frac{E-E_R + E_R \log ( E_R/E )}
{E_R-E + E \log (E /E_R)} .
 \notag 
\end{alignat}
This also gives the symmetric fluctuation about the uncorrelated
average in the individual components as $\Delta_{k2} =\langle
k_t^{(1)} \rangle -\langle k_t^{(22)} \rangle$ (see
Fig.~\ref{schemkt}).

Turning to the correlated emissions, due to the energy and angular
ordering in the correlated emission, this contribution is easily
evaluated and gives
\begin{alignat}{5}
\langle k_t^{(21)} \rangle_{\text{corr.}}
& = \frac{E}{P'_2} \frac{\as\;  C_F C_A}{\pi} \int_{\xi_R}^{\xi}\frac{d \xi'}{\xi'} 
\int_{E_R/E}^1 dz \, P(z)\, z \, \sqrt{\xi'} \,  
\int_{\xi_R}^{\xi'}\frac{d \xi''}{\xi''} 
\int_{E_R/E}^z dz' \, P(z') \notag \\
& =
\frac{4}{\log^2(E/E_R)\log^2(\xi/\xi_R)}
\left(E_R-E +E_R \log\left[ \frac{E}{E_R}\right]\right) \left(
2-2\sqrt{\xi/\xi_R} + \sqrt{\xi/\xi_R} \log\left[\frac{\xi}{\xi_R}\right]
\right),
\label{ave_two_uncorr}
\end{alignat}
and
\begin{alignat}{5}
\langle k_t^{(22)} \rangle_{\text{corr.}}
&= E\frac{E}{P'_2} \frac{\as\;  C_A C_F}{\pi}  \int_{\xi_R}^{\xi}\frac{d \xi'}{\xi'} 
\int_{E_R/E}^1 dz \, P(z) \,  
\int_{\xi_R}^{\xi'}\frac{d \xi''}{\xi''} 
\int_{E_R/E}^z dz' \, P(z') \, z' \, \sqrt{\xi''} \notag \\
&=\frac{4}{\log^2(E/E_R)\log^2(\xi/\xi_R)}
\left(E -E_R+E_R \log\left[ \frac{E_R}{E}\right]\right) \left(
2\sqrt{\xi/\xi_R} - 2 - 2 \log\left[\frac{\xi}{\xi_R}\right]
\right).
\label{ave_two_uncorr}
\end{alignat}
\begin{figure}[t]
\linethickness{1mm}
\line(1,0){300}
\put(0,-5){\line(0,1){10}}
\put(-300,-5){\line(0,1){10}}
\put(-318,12){$E_R\sqrt{\xi_R}$}
\put(-5,12){$E$}
\put(-200,12){$\langle k_t^{(1)} \rangle$}	
\put(-241,12){$\langle k_t^{(22)} \rangle_{\text{u}}$}	
\put(-165,12){$\langle k_t^{(21)} \rangle_{\text{u}}$}
\put(-115,12){$\langle k_t^{(21)} \rangle_{\text{c}}$}
\put(-280,12){$\langle k_t^{(22)} \rangle_{\text{c}}$}
\put(-216,-12){$\Delta_{k2}$}	
\put(-174,-12){$\Delta_{k2}$}	
\put(-57,12){\Large $\cdots$}	
\linethickness{.5mm}
\put(-265,-5){\line(0,1){10}}
\put(-230,-5){\line(0,1){10}}
\put(-190,-5){\line(0,1){10}}
\put(-150,-5){\line(0,1){10}}
\put(-100,-5){\line(0,1){10}}	
\caption{Schematic distribution for the average splitting scales
  $\langle k_t \rangle$ for $1$ and $2$ gluon emissions, in the latter
  case split in terms of the correlated and uncorrelated component.
  The scale $E$ represents the initial energy of the emitting parton.
\label{schemkt}
}
\end{figure}
\begin{figure}[t]\includegraphics[scale=0.60]{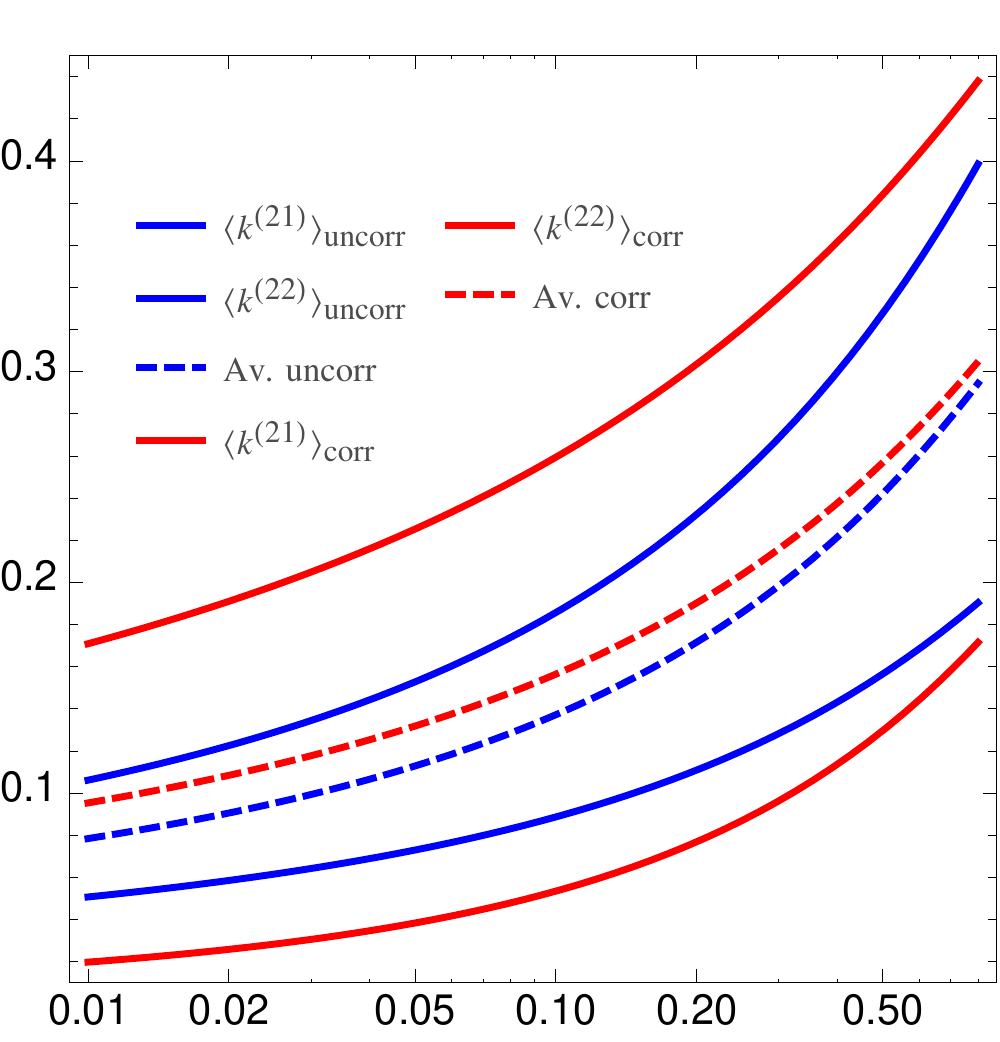}
\put(-210,112){$\dfrac{k_t}{E\sqrt{\xi_R}} $}	
\put(-44,-12){$R$}	
\hspace{1.5cm}
\includegraphics[scale=0.60]{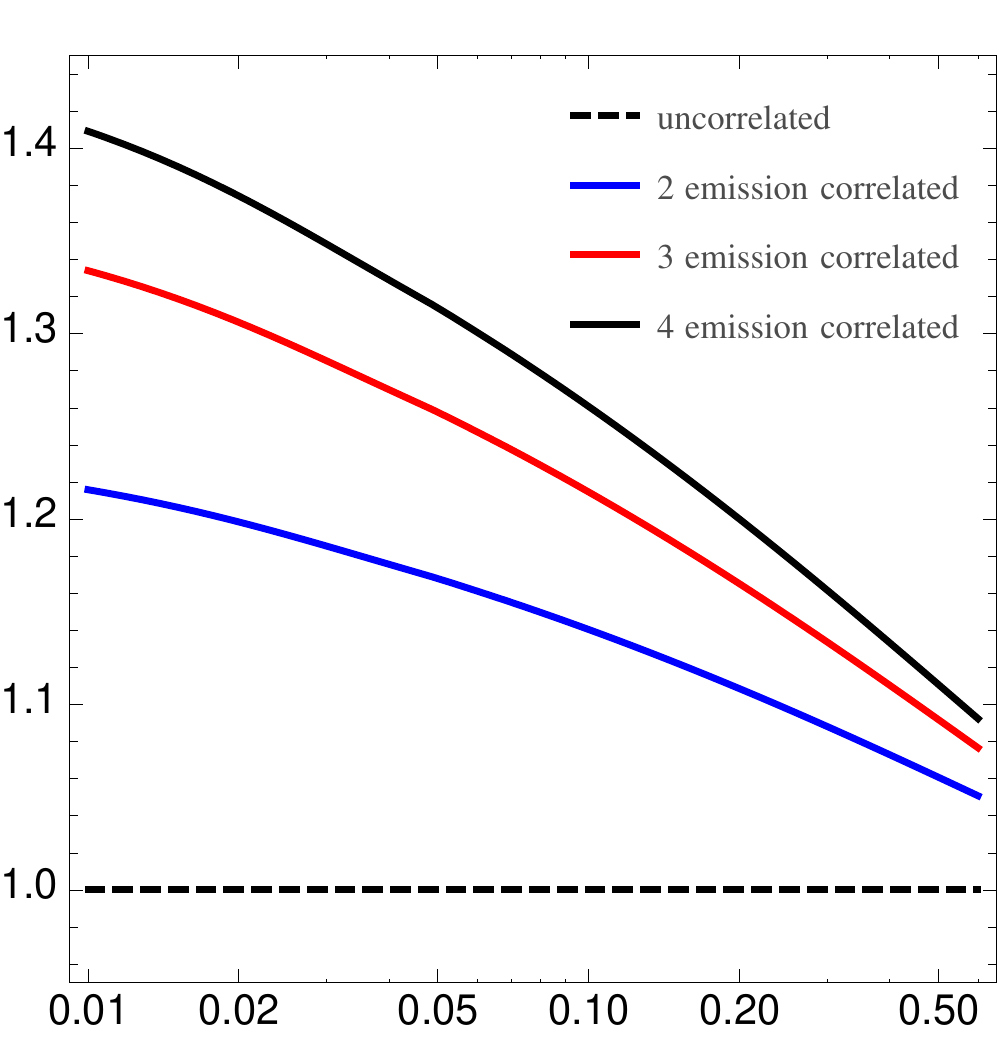}
\put(-210,112){ $\dfrac{\langle k_t^{(i)} \rangle}{\langle k_t^{(1)} \rangle}$}	
\put(-44,-12){$R$}	
\caption{(Left) Distribution for the scaled average splitting scale
  $\langle k_t \rangle/E$ for $2$ gluon emission as a function of $R$,
  divided among the correlated and uncorrelated components.  (Right)
  Ratio of the average splitting scale per jet for 1,2,3,4 emissions
  for the all correlated contribution. }
\label{splitscales234}
\end{figure}

\subsubsection*{Three emissions}
For the $3$ emission component we proceed as before though we have 4
splitting histories, which differ by their sequence of primary (P) and
secondary (S) emissions. They can be written $PPP$, $PPS$, $PSP$ and
$PSS$ where
\begin{equation*}
\text{PPP \;:} \quad \includegraphics[width=0.1\textwidth]{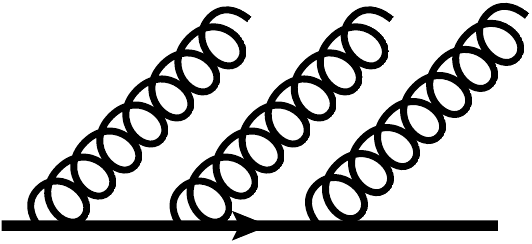}
\qquad \quad
\text{PPS\;:} \quad \includegraphics[width=0.1\textwidth]{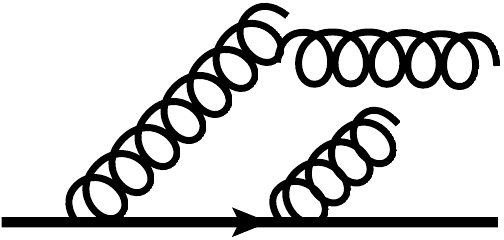}
\qquad \quad
\text{PSP\;:} \quad \includegraphics[width=0.1\textwidth]{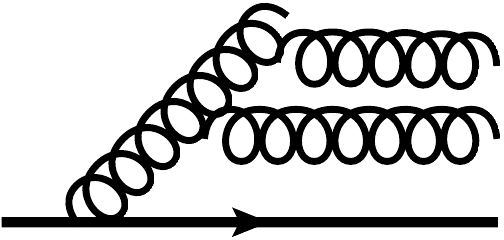}
\qquad \quad
\text{PSS\;:} \quad \includegraphics[width=0.1\textwidth]{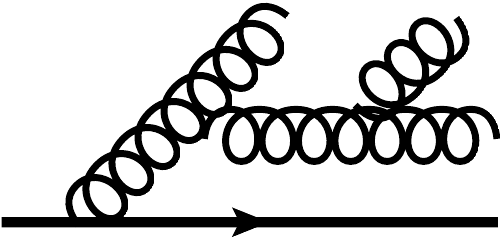}
\end{equation*}
First we have the normal Poisson relation that $\langle k_t^{(32)}
\rangle_{PPP} = \langle k_t^{(1)} \rangle $ and $\langle k_t^{(31)}
\rangle_{PPP} - \langle k_t^{(32)} \rangle_{PPP} = \langle k_t^{(32)}
\rangle_{PPP}-\langle k_t^{(33)} \rangle_{PPP} $ in the all correlated
part.  We give here only the analytic form of the $PSS$ (or fully
correlated) contribution for the hardest emission
\begin{alignat}{5}
\langle k_t^{(31)} \rangle_{\text{PSS}}
& = 
\frac{288}{\log^3(E/E_R)\log^3(\xi/\xi_R)E_R}
\left(E \log \left(\frac{E}{E_R}\right) (E_R-4 E) 
\log \left(\frac{E}{E_R} \right)+4 E-2 E_R)-2 E+2 E_R 
\right) \notag \\
& \quad \;\, \left( \sqrt{\xi/\xi_R} \left(  \frac{1}{8} \log ^2
\left(\frac{\xi}{\xi_R}\right)-\frac{1}{2} \log \left(\frac{\xi}{\xi_R}\right)
+1\right)- 1 \right) \, .
\label{ave_two_uncorr}
\end{alignat}

As with the 2 correlated emissions the average of the splitting scale
for 3 correlated emissions $(\langle k_t^{(31)}\rangle +\langle
k_t^{(32)} \rangle + \langle k_t^{(33)} \rangle_{PSS})/3$ is larger
than the single emission result $\langle k_t^{(1)}\rangle$.

\subsubsection*{Results for all multiplicity}

In principle we can repeat the procedure for an arbitrary splitting
history, and thus for all multiplicities.  However, the current method
is extremely tedious.  Instead we will focus only on the all
correlated contribution since it has the largest deviation from a
Poissonian splitting distribution for a fixed multiplicity.  In other
words, all average splitting scales for an arbitrary history fall
somewhere between the all correlated and the all primary results.  The
leading contribution to the all correlated average at fixed coupling
is obtained from the limit
\begin{alignat}{5}
\label{hard_int_solved}
\langle k_t^{(\infty)} \rangle =  \lim_{j\to \infty}
 \dfrac{\left(E\,  \prod_{k=1}^j 
\int_{\xi_R}^{\xi_{k+1}} \frac{d\xi_k}{\xi_k} \int_{E_R/E}^{z_{k+1}} \frac{dz_k}{z_k} \right) 
\sum_{k=1}^{j} z_k \sqrt{\xi_k/\xi_R}}
{j \left( \prod_{k=1}^j 
\int_{\xi_R}^{\xi_{k+1}} \frac{d\xi_k}{\xi_k} \int_{E_R/E}^{z_{k+1}} \frac{dz_k}{z_k} \right)}
\end{alignat}
which for the general case we are not able to take analytically.
However a trick at our disposal is the small $R$ limit, where the
small $R$ behaviour of the correlated plots in
Fig.~\ref{splitscales234} suggests that this converges to a finite
non-zero value. We find the following series representation for
\eq{hard_int_solved} divided by the single emission average in that
case,
\begin{alignat}{5}
\label{series_hint}
\frac{\langle k_t^{(\infty)} \rangle}{\langle k_t^{(1)}\rangle} =    \lim_{j\to \infty}
 \frac{(j+1)!}{(1-e)\log^j(e)}\left(1-e + \sum_{k=1}^{j} \frac{\log^{k}(e)}{k!}\right)
 \,\; + \,\; \mathcal{O}(1/\log(1/R)).
\end{alignat}
Substituting the asymptotic Stirling approximation for the the sum in 
brackets, we find the surprisingly simple expression
\begin{alignat}{5}
\label{series_hint_2}
\frac{\langle k_t^{(\infty)} \rangle}{\langle k_t^{(1)}\rangle} &=    \lim_{j\to \infty}
 \frac{(j+1)!}{(1-e)\log^j(e)}\left(-\frac{\log^{j+1}(e)}{j^{j+1} \exp({-j})\sqrt{2\pi j} }\right)
 \,\; + \,\; \mathcal{O}(1/\log(1/R))
 \notag \\
 & =
 \frac{\log(e)}{e-1}
\end{alignat}
This in turn implies the small $R$ behaviour of the correlated
emission scale $\langle k_t^{(\infty)} \rangle = E/\log(\sqrt{2}/R)$,
which can be compared to the pure Poisson single emission result at
small-$R$, given from~\eq{1lowor} by $\langle k_t^{(1)} \rangle =
E(1-e)/(\log(e)\log(\sqrt{2}/R))$ Note that these expression very
nicely enforce the intuitive fact that $\langle k_t^{(1)} \rangle <
\langle k_t^{(\infty)} \rangle $ for all values of $E_R$ and $R$, even
though the latter does not depend on $E_R$ explicitly.  The
correlated emission component, via the precise cancellation
in~\eq{series_hint} in the $R\ll 1$ limit loses memory of the scale
$E_R$.

\subsection{Discussion of results}

We summarise the findings of our calculations for the average $k_t$
splitting scales. For arbitrary multiplicity, the all uncorrelated
contribution produces a Poisson process where the average splitting
scale is a constant for all $n$.  The largest deviation from a Poisson
process is the all correlated emission component, which puts this
observable in a similar class as the average jet
multiplicity~\cite{Ellis:1991qj}, in the sense of being maximally
sensitive to correlated emissions.  We were able to show that at
leading order in the small-$R$ limit, the all-correlated contribution
converges to a constant.  Of course physically it must converge to
something since it is bounded from above and below, but what is
surprising is the simple analytic result in this case.  We speculate
that for the generic leading logarithmic QCD behaviour of the average
jet multiplicity it converges to a constant
\begin{equation}
\lim_{n\to \infty} \frac{\langle \bar{k}_t^{(n)}\rangle }
{  \langle \bar{k}_t^{(1)} \rangle } = 1+c,  
\end{equation}
with $c> 0$.  Whether or not this behaviour is realised in realistic
QCD environments, for example in the internal dynamics of large jets,
is a question which can be answered with a dedicated phenomenological
study.

\section{Conclusions}
\label{sec:conc}

In this paper we have studied jets at high multiplicity.  Using the
Gen-$k_t$ generating functional, we found that scaling patterns in the
rates emerged in particular limits in energy, angle and multiplicity.
We speculated that finite area effects are responsible for depressing
the geometric scaling in the jet ratios and attempted to quantify the
area distribution as accurately as possible.  Using these
considerations we managed to describe the area distribution coming
from a parton shower simulation.  The area depletion effect on the
rates on the other hand, was found to be significant but too small to
explain the tilt in the $n$-jets distribution at finite $R$,
indicating that additional terms are necessary to fully, analytically
describe the parton shower generated $n_{\text{jet}}$ distribution.
In the second part of this paper we looked into a new observable from
the perspective of scaling and found that it could also be shown,
formally at least, to produce a type of universal behaviour at high
multiplicity, which emerges due to the dominance of correlated
emissions in this limit.

We did not undertake a detailed phenomenological study in this work.
However, one can imagine that sub-jet multiplicities with an energy
cut sufficiently small might also show an extended regime of
generalised types of scaling.  Furthermore, the distribution of jet
areas can be extended to regimes where this quantity is useful, for
example assessing pile-up contributions and non-perturbative effects.
Finally, we speculate that additional distributions and observables
may show surprising emergent properties at high multiplicity, which
could lead to experimental and theoretical progress.

\newpage

\begin{center}
{\bf Acknowledgments}
\end{center}
We thank Steffen Schumann for helpful comments on the draft.  PS
acknowledges support from the european Union as part of the FP7 Marie
Curie Initial Training Network MCnet ITN (PITN-GA-2012-315877) and the
IMPRS for Precision Tests of Fundamental Symmetries.  EG acknowledges
support by the Bundesministerium f\"ur Bildung und Forschung under
contract 05H2012.
\begin{appendix}

\section{Exponentiated form of the evolution equation}
\label{app:evo}
We start from Eq.~(4.7) in~\cite{Gerwick:2012fw}, which is
\begin{align}
  \label{eq:evo_orig}
  \Phi_q(E,\xi) &= u + \int\limits^\xi_{\xi_R} \frac{d\xi'}{\xi'}
  \int\limits^1_{E_R/E} dz \frac{\alpha_s(k_T^2)}{2\pi}
  P_{q\rightarrow qg} \Phi_q(E,\xi') \left( \Phi_g(zE,\xi') -1 \right)
  \notag \\
  \Phi_g(E,\xi) &= u + \int\limits^\xi_{\xi_R} \frac{d\xi'}{\xi'}
  \int\limits^1_{E_R/E} dz \frac{\alpha_s(k_T^2)}{2\pi} \left[
  P_{g\rightarrow gg} \Phi_g(E,\xi') \left( \Phi_g(zE,\xi') -1 \right)
  \right. \notag \\ 
  &\quad \left. + P_{g\rightarrow q\bar{q}} \left( \Phi_q^2(E,\xi') - 
  \Phi_g(E,\xi')\right) \right].
\end{align}
Using $e=E/E_R$ we rewrite it in the form
\begin{align}
  \label{eq:evo_rewr}
  \Phi_i(e,\xi) &= u + \int\limits^\xi_{\xi_R} \frac{d\xi'}{\xi'}
  \int\limits^1_{1/e} dz \frac{\alpha_s(ze,\xi')}{2\pi}
  \sum\limits_{j,k}  P_{i\rightarrow jk}(z) \Phi_i(e,\xi')
  \left( \frac{ \Phi_j(e,\xi') \Phi_k(E(z),\xi') }{ \Phi_i(e,\xi') }
    -1 \right),
\end{align}
where the running coupling is expressed as
follows~\cite{Gerwick:2012fw}
\begin{align}
  \alpha_s(k_T^2) &= \frac{ \pi}{ b_0 \text{log} \frac{ z^2e^2E_R^2\xi}{
      \Lambda^2}}.
\end{align}
$i$ is either $q$ or $g$ and we sum over all allowed splittings.  Now
we take the derivative with respect to $\xi$ and find
\begin{align}
  \label{eq:evo_diff}
  \frac{d\Phi_i(e,\xi)}{d\xi} &= \frac{1}{\xi} \Phi_i(e,\xi)
  \int\limits^1_{1/e} dz \frac{\alpha_s(ze,\xi)}{2\pi}
  \sum\limits_{j,k}  P_{i\rightarrow jk}(z) \left( \frac{
      \Phi_j(e,\xi) \Phi_k(E(z),\xi) }{ \Phi_i(e,\xi) } -1 \right).
\end{align}
Taking into account that $\Phi_i(e,1)=u$ the general closed solution
is
\begin{align}
  \label{eq:evo_sol}
  \Phi_i(e,\xi) &= u \text{~exp} \left[ \int\limits^\xi_{\xi_R}
    \frac{d\xi'}{\xi'} \int\limits^1_{1/e} dz
    \frac{\alpha_s(ze,\xi')}{2\pi} \sum\limits_{j,k}
    P_{i\rightarrow jk}(z) \left( \frac{ \Phi_j(e,\xi)
        \Phi_k(E(z),\xi) }{ \Phi_i(e,\xi) } -1 \right)
    \right].
\end{align}

\section{Closed solution in the staircase limit}
\label{app:solve_stair}
We start from~\eq{eq:evo_expand}
\begin{align}
  \label{eq:evo_expand1}
  \Phi_g(e,\xi) &= u \text{~exp} \left( \int\limits^\xi_{\xi_R}
    \frac{d\xi'}{\xi'} \int\limits^1_{1/e} dz
    \frac{\alpha_s(z,\xi')}{2\pi} P_{g\rightarrow gg}(z) \left[
      \Phi_g(e,\xi') + \sum\limits^\infty_{n=1} \frac{(e(z-1))^n}{n!}
      \frac{ d^n \Phi_g(e,\xi')}{ de^n } -1 \right] \right).
\end{align}
Note that due its exponantial form all derivatives of $\Phi_g$ can be
written in the following form
\begin{align}
  \label{eq:phi_derivativ}
  \frac{ d^n \Phi_g(e,\xi')}{ de^n} &= \Phi_g(e,\xi') \times
  \mathcal{DP}[n](e,\xi';\Phi_g),
\end{align}
where $\mathcal{DP}[n]$ is a polynomal of inner derivatives of
$\Phi_g$. We plug this in eq.~(\ref{eq:evo_expand1}) and get
\begin{align}
  \label{eq:evo_expand2}
  \Phi_g(e,\xi) &= u \text{~exp} \left[ \int\limits^\xi_{\xi_R}
    \frac{d\xi'}{\xi'} \int\limits^1_{1/e} dz
    \frac{\alpha_s(z,\xi')}{2\pi} \right. \times \notag \\
  & \quad \left. P_{g\rightarrow gg}(z) \left( \Phi_g(e,\xi') +
      \Phi_g(e,\xi') \underbrace{ \sum\limits^\infty_{n=1}
        \frac{(e(z-1))^n}{n!}  \mathcal{DP}[n] (e,\xi';\Phi_g)
      }_{\mathcal{T}(z,e,\xi')} -1 \right) \right].
\end{align}
Up to this point we have not gained any new insight.  We can now
introduce more new symbolds to rewrite eq.~(\ref{eq:evo_expand2})
and bring it in the form
\begin{align}
  \label{eq:evo_expand3}
  \Phi_g(e,\xi) &= u \text{~exp} \left[ \int\limits^\xi_{\xi_R} d\xi'
    \left( \Phi_g(e,\xi') -1 \right)~ \underbrace{ \frac{1}{\xi'}
      \int\limits^1_{1/e} dz \frac{\alpha_s(z,\xi')}{2\pi}
      P_{g\rightarrow gg}(z)}_{\gamma_g(e,\xi')} + \right. \notag \\
  &\quad \left. \int\limits^\xi_{\xi_R} d\xi' \Phi_g(e,\xi')
    ~\underbrace{ \frac{1}{\xi'} \int\limits^1_{1/e} dz
      \frac{\alpha_s(z,\xi')}{2\pi} P_{g\rightarrow gg}(z)
      \mathcal{T}(z,e,\xi')}_{r(e,\xi')} \right] \notag \\
  &= u \text{~exp} \left[ \int\limits^\xi_{\xi_R} d\xi' \left(
      \Phi_g(e,\xi') -1 \right) \gamma_g(e,\xi') +
    \int\limits^\xi_{\xi_R} d\xi' \Phi_g(e,\xi') r(e,\xi') \right].
\end{align}
Taking the derivative this defines a differential equation of the form
\begin{align}
  \label{eq:evo_diff}
  \frac{d\Phi_g(e,\xi)}{d\xi} &= \Phi_g(e,\xi) \times \left[
    \gamma_g(e,\xi) \left( \Phi_g(e,\xi) -1 \right) + r(e,\xi)
    \Phi_g(e,\xi) \right] \notag \\ 
  \Phi_g(e,\xi_r) &= u.
\end{align}
We note that for the case $r(e,\xi)\rightarrow 0$
eq.~(\ref{eq:evo_diff}) produces exact staircase
scaling~\cite{Gerwick:2012hq}. Thus we connect this term to staircase
breaking. We also can solve eq.~(\ref{eq:evo_diff}) for arbitrary
$r(e,\xi)$. The solution is
\begin{align}
  \label{eq:evo_solve}
  \Phi_g(e,\xi) &= \frac{ 1 }{ 1 + \frac{ (1-u) }{ u \Delta_g(e,\xi) }
    - \underbrace{\int\limits^\xi_{\xi_R} d\xi' \frac{ \Delta_g(e,\xi')
      }{ \Delta_g(e,\xi)} r(e,\xi')}_{\mathcal{R}(e,\xi)} }
\end{align}
where $\Delta_g(e,\xi) = \text{exp} \left[ -\int\limits_{\xi_R}^\xi
  d\xi'\gamma(e,\xi') \right] $ is the Sudakov form factor. This is a
neat result. We find that we can write the generating functional in general in a
staircase like form together with an yet unspecified staircase scaling
breaking term. We check the solution explicitly by differentiating
eq.~(\ref{eq:evo_solve}) with respect to $\xi$. We note that we have
\begin{align}
  \label{eq:diffs1}
  \frac{ d \Delta_g(e,\xi) }{ d\xi} &= -\gamma(e,\xi) \Delta_g(e,\xi)
  \notag\\
  \frac{ d \mathcal{R} (e,\xi) }{d\xi} &= r(e,\xi) + \gamma(e,\xi)
  \mathcal{R} (e,\xi).
\end{align}
Therefore, we get
\begin{align}
  \label{eq:diffs2}
  \frac{d\Phi_g(e,\xi)}{d\xi} &= (-1) \left( 1 + \frac{1-u}{ u
      \Delta_g(e,\xi) } - \mathcal{R}(e,\xi) \right)^{-2} \notag \\
  &~\times \left[ \frac{1-u}{ u \Delta_g(e,\xi) } \gamma(e,\xi)
    \Delta_g(e,\xi) - r(e,\xi) - \gamma(e,\xi) \mathcal{R}(e,\xi)
  \right] \notag\\
  &= \Phi_g(e,\xi) \times \left[ \gamma_g(e,\xi) \left( \Phi_g(e,\xi)
      -1 \right) + r(e,\xi) \Phi_g(e,\xi) \right].
\end{align}
To evolve further we need to employ some assumptions. This means we
have to drop the explicit $\Phi_g$ dependence in $\mathcal{T}$. For
example we could expand around $e\approx 1$ and write $\Phi_g$ as an
explicit series in $u$, dropping all higher terms in $e$. To find the
leading $u$ dependence we plug in all previous definitions and find
that
\begin{align}
  \label{eq:full_R}
  \mathcal{R}(e,\xi) &= \int\limits^\xi_{\xi_R} d\xi' \frac{
    \Delta_g(e,\xi') }{ \Delta_g(e,\xi)} r(e,\xi') \notag \\
 &= \int\limits^\xi_{\xi_R} d\xi' \frac{ \Delta_g(e,\xi') }{
    \Delta_g(e,\xi)} \frac{1}{\xi'} \int\limits^1_{1/e} dz
  \frac{\alpha_s(z,\xi')}{2\pi} P_{g\rightarrow gg}(z)
  \mathcal{T}(z,e,\xi') \notag \\
  &= \int\limits^\xi_{\xi_R} d\xi' \frac{ \Delta_g(e,\xi') }{
    \Delta_g(e,\xi)} \frac{1}{\xi'} \int\limits^1_{1/e} dz
  \frac{\alpha_s(z,\xi')}{2\pi} P_{g\rightarrow gg}(z)
  \sum\limits^\infty_{n=1} \frac{(e(z-1))^n}{n!}  \mathcal{DP}[n]
  (e,\xi';\Phi_g).
\end{align}
The task at hand is to find a $\Phi_g$ independent approximation for
$\mathcal{DP}[n] $. An obvious step is to truncate the Tayler
expansion at $n=0$. We are then left with 
\begin{align}
  \label{eq:dp1}
  \mathcal{DP}[1] (e,\xi';\Phi_g) &= \frac{ d\Phi_g(e,\xi')
  }{\Phi_g(e,\xi') de }.
\end{align}
We need to find the significant part in the limit $e\rightarrow 1$ and
its $u$ dependence. We have
\begin{align}
  \label{eq:dp2}
  \mathcal{DP}[1] (e,\xi';\Phi_g)&=\int\limits^{\xi'}_{\xi_R}
  \frac{d\xi''}{\xi''} \left[ \frac{ \alpha_s(1/e,\xi'')
      P_{g\rightarrow gg}(1/e) } {2\pi~e^2} \left(
      \underbrace{\Phi_g(1,\xi'') }_{=u} -1 \right) \right. +
  \notag\\
  & \left. \int\limits^1_{1/e} dz \frac{ \alpha_s(z,\xi'') P_{g
        \rightarrow gg} (z) z}{ 2\pi} \left[
      \left. \frac{d\Phi(e,\xi'')}{de} \right|_{e=ze} \right] \right].
\end{align}
In principle we get a nested series of $\Phi_g$ differentiations. To
argue that first term is the most important we note that formally in
the $e\rightarrow 1$ limit we can write
\begin{align}
  \label{eq:dp3}
  \int\limits^1_{1/e} dz \frac{ \alpha_s(z,\xi'') P_{g \rightarrow gg}
    (z) z}{ 2\pi} \left[ \left. \frac{d\Phi(e,\xi'')}{de}
    \right|_{e=ze} \right] &\approx (1-1/e) \frac{ \alpha_s(1/e,\xi'')
    P_{g\rightarrow gg}(1/e) } {2\pi~e} \left[
    \left. \frac{d\Phi(e,\xi'')}{de} \right|_{e\approx 1} \right].
\end{align}
We note that the last term of eq.~(\ref{eq:dp3}) is exactly zero in
the formal limit $e\rightarrow 1$. This is our first argument that the
second term vanishes faster than the first one in the $e\rightarrow 1$
limit. We also can control its size this way and we set it to $u-1$
which is for sure over estimated as at the end of the day we have
$u\rightarrow 0$. We than see that for $e>1$ and $e\rightarrow 1$
\begin{align}
  \label{eq:dp4}
  1 &> e-1 \notag \\
  \frac{1}{e} &> \frac{ e-1}{e}\notag \\
  \frac{1}{e} &> (1- 1/e ).
\end{align}
Therefore, the first term is greater than the second one. However, this
is only valid in the $e\rightarrow 1$ limit. From this estimate we can
not deduce the uncertainty we introduce by dropping the second term
nor can we conclude on the range of $e$ where this estimate is valid,
because we do not know the correct $e$ dependence. Nevertheless, we
compute formally the $u$ dependence of the staircase breaking term in
the staircase limit $e\rightarrow 1$. Thus we write
\begin{align}
  \label{eq:dp5}
  \mathcal{DP}[1] (e,\xi';\Phi_g) &\approx (u-1)\int\limits^{\xi'}_{\xi_R}
  \frac{d\xi''}{\xi''} \frac{ \alpha_s(1/e,\xi'') P_{g\rightarrow
      gg}(1/e) } {2\pi~e^2} \left( u -1 \right) \notag \\
  &\equiv  \rho(e,\xi') \notag \\
  \mathcal{R}(e,\xi) &\approx (u-1)\int\limits^\xi_{\xi_R} d\xi' \frac{
    \Delta_g(e,\xi') }{ \Delta_g(e,\xi)} \frac{1}{\xi'}
  \int\limits^1_{1/e} dz \frac{\alpha_s(z,\xi')}{2\pi} P_{g\rightarrow
    gg}(z) (z-1) e (u-1) \rho(e,\xi') \notag \\
  &\equiv  \chi(e,\xi).
\end{align}

\end{appendix}


\bibliographystyle{ieeetr}
\bibliography{journal}

\end{document}

%% file: declare.tex
\def\as{\alpha_s}

\newcommand\one{\leavevmode\hbox{\small1\normalsize\kern-.33em1}}

\newcommand{\gev}{{\ensuremath\rm GeV}}

\def\slashchar#1{\setbox0=\hbox{$#1$}           
   \dimen0=\wd0                                 
   \setbox1=\hbox{/} \dimen1=\wd1               
   \ifdim\dimen0>\dimen1                        
      \rlap{\hbox to \dimen0{\hfil/\hfil}}      
      #1                                        
   \else                                        
      \rlap{\hbox to \dimen1{\hfil$#1$\hfil}}   
      /                                         
   \fi}

\renewcommand{\arraystretch}{1.3}